# Density, speed of sound, refractive index and relative permittivity of methanol, propan-1-ol or pentan-1-ol + aniline liquid mixtures. Application of the Kirkwood-Fröhlich model

Fernando Hevia*,[a], Víctor Alonso[b], Juan Antonio González[c], Luis Felipe Sanz[c], Isaías García de la Fuente[c], José Carlos Cobos[c]

[a] Université Clermont Auvergne, CNRS. Institut de Chimie de Clermont-Ferrand. F-63000, Clermont-Ferrand, France.

[b] Departamento de Física Aplicada. EIFAB. Campus Duques de Soria. Universidad de Valladolid. C/ Universidad s.n. 42004 Soria, Spain.

[c] G.E.T.E.F., Departamento de Física Aplicada. Facultad de Ciencias. Universidad de Valladolid. Paseo de Belén, 7, 47011 Valladolid, Spain.

*e-mail: fernando.hevia@termo.uva.es

# Abstract

Densities ($\rho$) and speeds of sound ($c$) at a temperature $T$ = 298.15 K, relative permittivities at 1 MHz ($\varepsilon_r$) and refractive indices at the sodium D-line ($n_D$) at $T$ = (293.15 K to 303.15 K), all of them at a pressure $p$ = 0.1 MPa, are reported for binary liquid mixtures alkan-1-ol + aniline. The alkan-1-ols considered are methanol, propan-1-ol and pentan-1-ol. Also, the values of the excess molar volume ($V_m^E$), excess isentropic compressibility ($\kappa_S^E$), excess speed of sound ($c^E$), excess refractive index ($n_D^E$), excess relative permittivity ($\varepsilon_r^E$) and its temperature derivative $(\partial \varepsilon_r^E / \partial T)_p$ are calculated and fitted to Redlich-Kister polynomials. The agreement among the reported data and other literature sources is analysed by comparing $V_m^E$, $n_D^E$, $\varepsilon_r^E$ and the deviation of $c$ from mole-fraction linearity ($\Delta c$). The positive excess molar internal energies at constant volume ($U_{m,V}^E$) show the dominance of the breaking of interactions between like molecules in the energy balance on mixing, particularly the breaking of strong dipolar interactions between aniline molecules. This contribution is also dominant for the $\varepsilon_r^E$ values, as they are negative and decrease with the length of the alkan-1-ol chain. Calculations on the concentration-concentration structure factor are consistent with these statements, revealing homocoordination in the studied systems. The $V_m^E$ are negative, which together with the positive $U_{m,V}^E$ indicate the existence of important structural effects in the studied mixtures. The application of the Kirkwood-Fröhlich model shows that the average relative orientation of neighbouring dipoles is similar in the mixtures methanol + aniline or + pyridine, in spite of the different character of the predominant interactions in the latter mixture (heterocoordination).

# 1. Introduction

The high polarizability of the aromatic ring, together with the $NH_2$ group, gives the aniline molecule very interesting properties. Thus, liquid mixtures of aniline and alkanes show high upper critical solution temperatures (UCSTs), as it is shown by the high value of this quantity for the heptane system, 343.11 K [1], which reveals very strong dipolar interactions between aniline molecules. Alkan-1-ol + aniline liquid mixtures are characterized by positive deviations from Raoult's law, as it is demonstrated by the following $G_m^E$/J·mol$^{-1}$ values at equimolar composition for the solutions with methanol (515, $T$ = 293.15 K) [2] or ethanol (644, $T$ = 313.15 K) [2]. It is to be noted that systems containing an alkan-1-ol and one primary, or secondary or tertiary aliphatic amine behave in the opposite way. For example, in the case of propan-1-ol + propan-1-amine system at equimolar composition and $T$ = 302.95 K, $G_m^E$ = –752 J·mol$^{-1}$ [3].

As a part of our research on liquid mixtures of alcohols and amines [4-15], we are engaged in a systematic study of the dielectric behaviour of alkan-1-ol + amine liquid mixtures by means of relative permittivities ($\varepsilon_r$), refractive indices at the sodium D-line ($n_D$), and the application of the Kirkwood-Fröhlich model by the calculation of the Kirkwood correlation factor [16,17], which also requires the knowledge of density ($\rho$) data. So far, we have reported measurements for systems containing cyclohexanamine [18,19], hexan-1-amine [13], *N*-propylpropan-1-amine [14] and *N,N*-diethylethanamine [15]. Now we aim to investigate the impact on the dielectric properties of the replacement of the mentioned aliphatic amines by aniline, as the thermodynamic behaviour of these mixtures is quite different (see above). Therefore, we provide measurements of $\varepsilon_r$ at 1 MHz frequency and $n_D$ over the temperature range (293.15 to 303.15) K and, complementarily, of $\rho$ and the speed of sound (*c*) at 298.15 K, for the liquid mixtures methanol, propan-1-ol or pentan-1-ol + aniline at 0.1 MPa.

A survey of literature data using Web of Science and the Thermolit machine shows that different research groups have provided results of their measurements of these properties [20-38]. We perform a critical comparative analysis of these data, which reveals the poor precision of some of them and, more importantly, large discrepancies that make it difficult to judge their accuracy. The measurements reported in this work will, hopefully, help to reveal which of the literature datasets are more likely to be accurate.

## 2. Experimental

### 2.1. Materials

Information about the pure compounds employed in the experiments is shown in Table 1. They were used without further purification. The measured thermophysical properties at the working temperatures and their dipole moments ($\mu$) are collected in Table 2. Comparison with literature values reveals a good agreement.

### 2.2. Apparatus and procedure

The mixtures were prepared in glass vessels of about 10 cm$^3$, and the concentration was measured by mass determination using an analytical balance Sartorius MSU125p, with a standard uncertainty of 5·10$^{-5}$ g. The weighing was corrected taking into account buoyancy effects. The standard uncertainty in the mole fraction is 0.0005. Molar quantities were calculated using the relative atomic mass Table of 2015 issued by the Commission on Isotopic Abundances and Atomic Weights (IUPAC) [39]. Pure liquids were stored with 4 Å molecular sieves (except methanol, because measurements were affected) in order to minimize the effects of the interaction with air components. The measurement cells (see below) were appropriately filled with the samples and closed to avoid liquid evaporation. The density of the pure compounds was measured along the experiments, remaining constant within the experimental uncertainty.

Temperatures were measured using Pt-100 resistances. They were calibrated using the triple point of water and the melting point of Ga, according to the ITS-90 scale of temperature, as reference points. The standard uncertainty of this quantity is 0.01 K for $\rho$ and $c$ measurements, and 0.02 K for $\varepsilon_r$ and $n_D$ measurements.

Densities and speeds of sound of the liquids were obtained by means of a vibrating-tube densimeter and sound analyser, Anton Paar model DSA 5000, with a temperature stability of 0.001 K. A description of the calibration procedure has been given elsewhere [40]. The relative standard uncertainty of the $\rho$ measurements is 0.0008. The determination of the speed of sound is based on the measurement of the propagation time of short acoustic pulses of 3 MHz centre frequency [41], which are repeatedly transmitted through the sample. The standard uncertainty of the $c$ measurements is 0.2 m·s$^{-1}$. The experimental technique was checked through the determination of the excess molar volume, $V_m^E$, and excess isentropic compressibility, $\kappa_S^E$ (see equations below), of the cyclohexane + benzene mixture at $T$ = (293.15 to 303.15) K, and there is good agreement between our results and published values [42-45].

The quantity $n_D$ was determined by means of a Bellingham + Stanley RFM970 automatic refractometer, which determines the critical angle at the wavelength of the sodium D-line (589.3 nm) and relates it to the corresponding refractive index. The standard uncertainty of $n_D$ is 0.00008. The temperature stability (0.02 K) is achieved by means of Peltier modules. The apparatus was calibrated following the recommendations by Marsh [46], using 2,2,4-trimethylpentane and toluene at $T =$ (293.15 to 303.15) K.

The experimental device to determine $\varepsilon_r$ and its calibration has been described in detail elsewhere [47]. It is based on precise impedance measurement by the auto-balancing bridge method in 4TP (Four-Terminal Pair) configuration, using a precision impedance analyser 4294A and a 16048G test lead from Agilent. A sample volume of $\approx$ 4.8 cm$^3$ fills a 16452A cell (parallel-plate capacitor, also from Agilent), which is immersed in a thermostatic bath LAUDA RE304, with a temperature stability of 0.02 K. The repeatability of the $\varepsilon_r$ measurements is 0.0001. The relative standard uncertainty of $\varepsilon_r$ was estimated to be 0.003 from the differences between our data and values available in the literature, in the range of temperature (288.15 to 333.15) K, for the following pure liquids: water, benzene, cyclohexane, hexane, nonane, decane, dimethyl carbonate, diethyl carbonate, methanol, propan-1-ol, pentan-1-ol, hexan-1-ol, heptan-1-ol, octan-1-ol, nonan-1-ol and decan-1-ol.

## 3. Equations

Assuming negligible dispersion and absorption of the acoustic wave in the liquid, the isentropic compressibility ($\kappa_S$) can be determined from $\rho$ and $c$ measurements using the Newton-Laplace equation:

$$\kappa_S = \frac{1}{\rho c^2} \qquad (1)$$

The isothermal compressibility ($\kappa_T$) is related to the molar volume ($V_m$), the molar isobaric heat capacity ($C_{pm}$), the thermal expansion coefficient ($\alpha_p$) and $\kappa_S$ through:

$$\kappa_T = \kappa_S + \frac{TV_m(\alpha_p)^2}{C_{pm}} \qquad (2)$$

The values $F^{id}$ of the thermodynamic properties of an ideal mixture at the same temperature and pressure as the real mixture are calculated from the well-established equations from Benson and Kiyohara [48-50]:

$$F^{id} = x_1 F_1^* + x_2 F_2^* \qquad (F = V_m, C_{pm}) \qquad (3)$$

$$F^{id} = \phi_1 F_1^* + \phi_2 F_2^* \qquad (F = \alpha_p, \kappa_T) \qquad (4)$$

where $F_i^*$ is the value of the property $F$ of pure component $i$, $x_i$ denotes the mole fraction of component $i$ and $\phi_i = x_i V_{mi}^* / V_m^{id}$ represents the volume fraction of component $i$. Ideal values of $\kappa_S$ and $c$ are calculated from the formulae [48]:

$$\kappa_S^{id} = \kappa_T^{id} - \frac{T V_m^{id} (\alpha_p^{id})^2}{C_{pm}^{id}} \qquad (5)$$

$$c^{id} = \left( \frac{1}{\rho^{id} \kappa_S^{id}} \right)^{1/2} \qquad (6)$$

being $\rho^{id} = (x_1 M_1 + x_2 M_2)/V_m^{id}$ ($M_i$, molar mass of the $i$ component).

The ideal dielectric and refractive properties are given by [51,52]:

$$\varepsilon_r^{id} = \phi_1 \varepsilon_{r1}^* + \phi_2 \varepsilon_{r2}^* \qquad (7)$$

$$n_D^{id} = \left[ \phi_1 (n_{D1}^*)^2 + \phi_2 (n_{D2}^*)^2 \right]^{1/2} \qquad (8)$$

$$\left[ \left( \frac{\partial \varepsilon_r}{\partial T} \right)_p \right]^{id} = \left( \frac{\partial \varepsilon_r^{id}}{\partial T} \right)_p \qquad (9)$$

Lastly, the excess properties are obtained using the relation:

$$F^E = F - F^{id} \qquad (10)$$

## 4. Results

Values of $\alpha_p = -(1/\rho)(\partial \rho/\partial T)_p$ of the pure compounds at $T = 298.15$ K and $p = 0.1$ MPa were obtained from linear regressions of experimental $\rho$ values in the range (293.15 to 303.15) K. The temperature derivative $(\partial \varepsilon_r^E / \partial T)_p = [(\partial \varepsilon_r / \partial T)_p]^E = (\partial \varepsilon_r / \partial T)_p - (\partial \varepsilon_r^{id} / \partial T)_p$ was calculated in an analogous fashion.

Experimental values of $\rho$, $c$, $V_m^E$, $\kappa_S^E$ and $c^E$ of alkan-1-ol (1) + aniline (2) liquid mixtures at $T = 298.15$ K and $p = 0.1$ MPa are collected as functions of $x_1$ (mole fraction of the alkan-1-ol) in Table 3. Table 4 collects their $\phi_1$, $\varepsilon_r$ and $\varepsilon_r^E$ values of as functions of $x_1$ in the $T$ range (293.15 to 303.15) K and $p = 0.1$ MPa, while Table 5 contains the experimental $x_1$, $\phi_1$, $n_D$ and $n_D^E$ values at the same conditions. The data of $(\partial \varepsilon_r^E/\partial T)_p$ at $T = 298.15$ K and $p = 0.1$ MPa are collected in Table 6.

The $F^E$ data were fitted by unweighted linear least-squares regressions to Redlich-Kister polynomials [53]

$$F^E = x_1(1-x_1)\sum_{i=0}^{k-1} A_i (2x_1 - 1)^i \qquad (11)$$

The number, $k$, of appropriate coefficients for each system, property and temperature has been determined by the application of an F-test of additional term [54] at a 99.5% confidence level. Table 7 includes the parameters $A_i$ obtained, and the standard deviations, $\sigma(F^E)$, defined by:

$$\sigma(F^E) = \left[ \frac{1}{N-k} \sum_{j=1}^{N} (F_{\text{cal},j}^E - F_{\text{exp},j}^E)^2 \right]^{1/2} \qquad (12)$$

where $j$ indexes the $N$ experimental data $F_{\text{exp},j}^E$, and $F_{\text{cal},j}^E$ is the corresponding value of the excess property $F^E$ calculated from equation (11).

The values of $V_m^E$, $\kappa_S^E$ and $c^E$ versus $x_1$, and those of $\varepsilon_r^E$, $(\partial \varepsilon_r^E/\partial T)_p$ and $n_D^E$ versus $\phi_1$, at $T = 298.15$ K and $p = 0.1$ MPa, are plotted together with their corresponding Redlich-Kister regressions in Figures 1-6.

### 4.1. Comparison with literature data

Data available in the literature on $\rho$, $c$, $\varepsilon_r$ and $n_D$ of alkan-1-ol + aniline liquid mixtures [20-38] are analysed and compared with our measurements in Table 8 and Figures 1, 4, 5 and S1-S8 (supplementary material). For comparison purposes, the deviation from mole-fraction linearity of the speed of sound, $\Delta c$, has been computed:

$$\Delta c = c - x_1 c_1 - x_2 c_2 \qquad (13)$$

From now on, we will denote by $n$ the number of carbon atoms of the alkan-1-ol, and by $n$OH the corresponding alkan-1-ol. We will now analyse the reliability of the data available in the literature on $V_m^E$, $\Delta c$, $n_D^E$ and $\varepsilon_r^E$ for the studied systems.

**a) Density**. As can be seen from Figures 1 and S1-S3, there are several sources of $V_m^E$ values that agree well with our data for the systems with 1OH [21,22], 3OH [31] or 5OH [30]. Furthermore, references [23,33] regarding the 4OH mixture are in accordance with our $V_m^E$ variation with $n$, and are very close to each other (Figure 1). From these two, only reference [23] reports data with the same symmetry as our measurements. We are evaluating the agreement of the data taking into account that the temperature difference between sources is not large. Data from reference [34] seem, therefore, unreliable, as they give large and positive $V_m^E$ values, in contrast with the large negative values from our data and from references [21,22]. Reference [38] presents also abnormally high $V_m^E$ values, which are even large and positive for *n* = 4,5,6; it can be then concluded that the data from this reference have to be taken with caution. The same can be said about the data from reference [35] regarding the 3OH mixture, as their values are similar to those of reference [38]. As for the temperature dependence of $V_m^E$, none of the references provides enough precision to confirm any tendency. For the case of 4OH mixtures, the available references [23,33] seem to show opposite trends.

**b) Speed of sound**. There are few references reporting *c* values of the systems under study. $\Delta c$ values calculated from reference [22] for the 1OH mixture agree well with our data (Figure S4), like they also do for $V_m^E$ (see above). Reference [36] leads to $\Delta c$ values for the 3OH mixture comparable to ours (Figure S5). Data from reference [38] do not agree well with our data (Figure S6), but this discordance was already observed on the very deviated values it gives for $V_m^E$.

**c) Refractive index**. Sources of $n_D$ data are even scarcer. For the 4OH mixture, $n_D^E$ values can be calculated using data from reference [23] (Figure 5) and, as it happened with $V_m^E$ values from this reference, they agree rather well with our $n_D^E$ variation with *n* (taking into account the large data scattering). Data from reference [21] for the 1OH mixture is not precise enough to calculate $n_D^E$, giving a huge data scattering (Figure 5). $n_D^E$ from reference [37] disagree with our values (Figure 5). Given the lack of a sufficient amount of reliable references, the objective conclusion is to mark all these $n_D$ sources as uncertain ([23,37]) or not precise enough for the calculation of $n_D^E$ ([21]).

**d) Relative permittivity**. There are more sources of $\varepsilon_r$ data than in the case of $n_D$, but they differ a lot among them and/or present a significant data scattering. For the 1OH mixture, none of the $\varepsilon_r$ sources leads to $\varepsilon_r^E$ values in agreement with ours (Figure S7). Nevertheless, it must be noticed that either the number of experimental points is rather low [20] or the data scattering is important [26]. We see, however, rather good agreement in $\varepsilon_r^E$ of the 3OH mixture for references [24,32], although with significant data scattering (Figures 4 and S8); there is no agreement in the case of reference [28], but the data scattering in this case is enormous (Figure

S8). The rest of the references [25,27,29] do not provide the $\varepsilon_r$ values for the pure compounds, and consequently $\varepsilon_r^E$ cannot be evaluated consistently.

## 5. Discussion

In this section, if nothing else is specified, the values of the thermophysical properties are given at $T = 298.15$ K and $x_1 = 0.5$, except for dielectric properties, which will be given at $\phi_1 = 0.5$. As before, we will denote by $n$ the number of carbon atoms of the alkan-1-ols and by $n$OH the alkan-1-ol with $n$ carbon atoms.

### 5.1. Excess molar volumes, molar enthalpies and isentropic compressibilities

$n$OH + aniline liquid mixtures show positive, or small and negative, $H_m^E$/J·mol$^{-1}$ values: –175 (1OH [55]), 351 (2OH [55]), 776 (3OH [55]), 944 (4OH [56]), 1011 (5OH [55]). They show an important positive contribution to this quantity coming from the rupture of interactions between molecules of the same species. In contrast, their $V_m^E$/cm$^3$·mol$^{-1}$ are negative: –0.902 (1OH, this work), –0.593 (3OH, this work), –0.398 (4OH [23]), –0.241 (5OH, this work). Positive $H_m^E$ values occurring together with negative $V_m^E$ values indicate the presence of relevant structural effects. In fact, it is possible to evaluate the magnitude of these effects on the $H_m^E$ results, and subtract them by performing the calculation of the excess molar internal energy at constant volume ($U_{m,V}^E$) [57]:

$$U_{m,V}^E = H_m^E - T \frac{\alpha_p}{\kappa_T} V_m^E \qquad (14)$$

Due to the lack of reliable data, in the application of equation (14) to $n$OH + aniline mixtures both $\alpha_p$ and $\kappa_T$ were considered ideal. Also, for pure 4OH $\alpha_p$ and $\kappa_T$ were taken from Riddick *et al.* [58]. This gives a very reasonable approximation of the corresponding $U_{m,V}^E$/J·mol$^{-1}$ values: 143 (1OH), 997 (3OH), 1095 (4OH), 1104 (5OH). We observe an important contribution to $H_m^E$ coming from the equation-of-state term (the second term in equation (14)). The $U_{m,V}^E$ results are positive, revealing that, in the balance of interactional effects, there is a dominance of the rupture of interactions between like molecules, even for the 1OH mixture, which shows a positive $U_{m,V}^E$. In this balance, the breaking of the dipolar interactions between aniline molecules plays a major role. This can be seen by comparing with the largely negative $H_m^E$ values for the $n$OH + hexan-1-amine mixtures [59]. The importance of dipolar interactions and structural effects in $n$OH + aniline mixtures is also supported by the poor results obtained for these systems in the framework of the ERAS model [60]. It can also be observed that, as *n* takes higher values, the increment of $H_m^E$, or $U_{m,V}^E$, when *n* is incremented by 1 unit tends to decrease. This can be ascribed to a lower and weaker association of longer alkan-1-ols.

Negative values of $\kappa_S^E$ are typical of solutions where there are relevant interactions between unlike molecules and/or structural effects [61,62]. $\kappa_S^E$/TPa$^{-1}$ values of *n*OH + aniline liquid mixtures are moderately negative: –109.7 (1OH), –72.5 (3OH), –37.3 (5OH), and their change with *n* is parallel to that of $V_m^E$. They can be interpreted in terms of a decrease of the free volume available for the molecules due to the structural effects that make $V_m^E$ negative, which make the mixtures less compressible. These values are consistent with the positive $c^E$/m·s$^{-1}$ values: 139.0 (1OH), 86.5 (3OH), 44.8 (5OH).

## 5.2. Excess relative permittivities

The value of $\varepsilon_r^E$ is affected negatively when interactions between like molecules are disrupted, whereas it is impacted either positively or negatively when interactions between unlike molecules are created, depending on the character of the formed multimers in terms of their effective response to an electric field. In mixtures of the type *n*OH + benzene there is a dominance of the rupture of interactions along mixing, showing therefore negative values of $\varepsilon_r^E$ (at 293.2 K [58,63]): –1.28 (1OH), –2.15 (3OH), –2.78 (4OH), –2.83 (6OH), –2.23 (8OH). As *n* increases, $\varepsilon_r^E$ decreases and then increases again, which is a typical behaviour of *n*OH + alkane (e.g. heptane [12,58,64,65]) or + amine (e.g. hexan-1-amine [13], propylpropan-1-amine [14], *N,N*-diethylethanamine [15], cyclohexanamine [18,19]), and can be attributed to the lower and weaker association of longer alkan-1-ols. The values of $\varepsilon_r^E$ in *n*OH + aniline systems are also negative: –0.778 (1OH), –1.850 (3OH), –2.082 (5OH), indicating a predominance of the rupture of *n*OH-*n*OH and aniline-aniline interactions. The aniline molecule can be considered as a benzene molecule to which the –NH$_2$ group is added, so a larger negative contribution to $\varepsilon_r^E$ from the rupture of interactions between like molecules is expected in comparison with *n*OH + benzene mixtures. Consequently, the higher $\varepsilon_r^E$ values presented by *n*OH + aniline suggest that the contribution of the formation of interactions between unlike molecules is positive, in such a way that the created multimers respond more effectively to the electric field than the structures that existed in the pure liquids.

$\varepsilon_r^E$ decreases when *T* is increased for the 1OH system, and increase in line with *T* for the mixtures with 3OH or 5OH. A similar trend is encountered when aniline is replaced by hexan-1-amine. The negative $(\partial \varepsilon_r^E / \partial T)_p$ results for the solution containing 1OH may indicate that, at higher temperatures, the multimers in the mixture become of smaller size and/or cyclic nature and that this effect predominates over those related to the breaking of the interactions between like molecules, which should contribute positively to $(\partial \varepsilon_r^E / \partial T)_p$ as they decrease with $T$. Longer alkan-1-ols are more weakly self-associated and the permittivity of the ideal state decreases with the increasing of temperature more rapidly than $\varepsilon_r$ in such a way that $(\partial \varepsilon_r^E / \partial T)_p$ values become positive.

Pyridine shows a higher dipole moment (2.21 D [66]) and relative permittivity (12.91 [58]). It is interesting that, nevertheless, the strength of amine-amine interactions in pure pyridine is lower than in pure aniline, as revealed by the upper critical solution temperatures of their mixtures with heptane: 343.11 K (aniline [1]) > 255.2 K (pyridine [67]). In contrast with aniline, the liquid mixture 1OH + pyridine shows a positive $\varepsilon_r^E$ value (2.85 [58,68]), indicating positive contributions from the formed multimers, which are predominant over the negative contributions from the rupture of interactions. This is consistent with the fact that $H_m^E$(1OH + pyridine) = –707 J·mol$^{-1}$ [69] is somewhat lower than $H_m^E$(1OH + aniline) (see above), and suggests that strong 1OH-pyridine interactions could be enhancing the dipole moment of pyridine molecules in the mixed state. It is interesting to note that the results from $\varepsilon_r^E$ are also in accordance with those from the concentration-concentration structure factor ($S_{CC}(0)$) formalism. The concentration-concentration structure factor, defined by [70,71]:

$$S_{CC}(0) = \frac{RT}{\left(\frac{\partial^2 \Delta G_m}{\partial x_1^2}\right)_{T,p}} = \frac{x_1 x_2}{1 + \frac{x_1 x_2}{RT}\left(\frac{\partial^2 G_m^E}{\partial x_1^2}\right)_{T,p}} > 0 \qquad (15)$$

where $\Delta G_m$ is the molar Gibbs energy of mixing, is a way to study fluctuations in a binary mixture. For ideal mixtures, $S_{CC}(0) = x_1 x_2$. If $S_{CC}(0) > x_1 x_2$, the dominant trend in the system is homocoordination (separation of the components), whereas if $S_{CC}(0) < x_1 x_2$, the dominant trend in the solution is heterocoordination (compound formation). $S_{CC}(0)$ values have been evaluated for some $n$OH + amine systems by means of the DISQUAC model [3,68]. For the aniline mixtures [60], $S_{CC}(0)$ values are higher than the ideal ones: 0.398 (1OH, $T$ = 293.15 K), 0.461 (3OH), 0.493 (4OH), 0.504 (5OH). They are, therefore, characterized by homocoordination. In contrast, pyridine solutions [72] show $S_{CC}(0)$ values that deviate negatively from ideality: 0.235 (1OH), 0.230 (3OH, $T$ = 313.15 K), 0.230 (4OH, $T$ = 313.15 K). Consequently, these mixtures are characterized by heterocoordination.

## 5.3. Molar refraction

The molar refraction, $R_\text{m}$, is proportional to the average electronic contribution to the polarizability, $\alpha_\text{e}$, from one molecule in a macroscopic sphere of liquid [16,17], and consequently it is a measure of dispersive interactions [17,73]. It is defined by the so-called Lorentz-Lorenz equation [16,17]:

$$R_\text{m} = \frac{n_\text{D}^2 - 1}{n_\text{D}^2 + 2} V_\text{m} = \frac{N_\text{A} \alpha_\text{e}}{3\varepsilon_0} \qquad (16)$$

where $N_\text{A}$ is Avogadro's constant and $\varepsilon_0$ the vacuum permittivity. The $R_\text{m}$/cm$^3$·mol$^{-1}$ values for $n$OH + aniline mixtures (at equimolar composition) reveal more important dispersive interactions for larger $n$: 19.5 (1OH), 24.2 (3OH), 28.9 (5OH). The corresponding excess values, $R_\text{m}^\text{E} = R_\text{m} - R_\text{m}^\text{id}$, can be calculated substituting ideal values in equation (16) for $R_\text{m}^\text{id}$. For $n$OH + aniline systems, the $R_\text{m}^\text{E}$/cm$^3$·mol$^{-1}$ values at equimolar composition are negative: –0.27 (1OH), –0.17 (3OH), –0.07 (5OH). They mean there is a loss in dispersive interactions along mixing with respect to the ideal state, in which dipoles of different components do not interact. The lower $R_\text{m}^\text{E}$ value encountered for the methanol solution may be due, in this case, to a formation of a larger number of hydrogen bonds between the components upon mixing. A similar trend is observed for alkan-1-ol + hexan-1-amine mixtures [13].

## 5.4. Kirkwood-Fröhlich model

The Kirkwood-Fröhlich model for dielectrics is based on fluctuation relations at zero field. It is assumed that the molecules are in a spherical cavity of an infinitely large dielectric. The high-frequency permittivity, $\varepsilon_\text{r}^\infty$, with only induced polarizability contributions, is used to treat macroscopically the induced polarizability. There is a local field in the cavity that plays the role of long-range dipolar interactions, and is calculated considering the outside of the cavity as a continuous medium of permittivity $\varepsilon_\text{r}$. The Kirkwood correlation factor, $g_\text{K}$, represents the effect of short-range interactions and informs about the deviations of the relative orientation of a dipole with respect to its neighbors from randomness. For a one-fluid model of a mixture of polar liquids [74], $g_\text{K}$ can be determined using [16,17,74,75]:

$$g_\text{K} = \frac{9 k_\text{B} T V_\text{m} \varepsilon_0 (\varepsilon_\text{r} - \varepsilon_\text{r}^\infty)(2\varepsilon_\text{r} + \varepsilon_\text{r}^\infty)}{N_\text{A} \mu^2 \varepsilon_\text{r} (\varepsilon_\text{r}^\infty + 2)^2} \qquad (17)$$

Here, $k_\text{B}$ is Boltzmann's constant; $N_\text{A}$, Avogadro's constant; $\varepsilon_0$, the vacuum permittivity; and $V_\text{m}$, the molar volume of the liquid at the working temperature, $T$. For polar compounds, $\varepsilon_\text{r}^\infty$ has been estimated using $\varepsilon_r^\infty = 1.1 n_D^2$ [76]. $\mu$ represents the dipole moment of the solution, estimated from the equation [74]:

$$\mu^2 = x_1 \mu_1^2 + x_2 \mu_2^2 \qquad (18)$$

where $\mu_i$ stands for the dipole moment of component $i$ (= 1,2). Calculations have been performed using smoothed values of $V_m^E$, $n_D^E$ and $\varepsilon_r^E$ at $\Delta x_1 = 0.01$. The source and values of $\mu_i$ used for alkan-1-ol + aniline mixtures are collected in Table 2.

The values of $g_K$ ($n$OH+aniline) obtained in this work are moderately high and decrease with $n$ (Figure 7): 1.94 (1OH), 1.72 (3OH). 1.50 (5OH). The $n$ variation can be ascribed to the lower and weaker solvation and self-association of the alkan-1-ol as its chain becomes longer. We note that $g_K$ (1OH + aniline) is considerably lower than $g_K$ (1OH + hexan-1-amine) [13] (Figure 7), indicating that solvation and the self-association of the alkan-1-ol, a key feature of 1OH + hexan-1-amine, is less important in the aniline mixture. The dependence of $g_K$ (1OH + aniline) with $\phi_1$ suggests that aniline is a better breaker of the 1OH network than hexan-1-amine. For the sake of comparison, we also show $g_K$ results when aniline is replaced by benzonitrile (293.15 K, [20]), a strongly polar, non-associated compound. In fact, in this case the $g_K$ curve increases even slowly with $\phi_1$ (Figure 7), as a result of a quite effective rupture of the self-association of 1OH by benzonitrile [77]. On the other hand, if aniline is kept but 1OH is replaced by a strongly polar, non-associated compound such as *N,N*-dimethylformamide [78], the values of $g_K$ are very low (Figure 7) as the high contribution to parallel alignment from the association of 1OH disappears.

Lastly, let us see that the $g_K$ curve of the system 1OH + pyridine [68] (Figure 7) is very similar to that of the mixture 1OH + aniline. As has been noted above, the former is characterized by homocoordination, whereas the latter by heterocoordination. The results from the model suggest that, even though the kind of interactions predominant in each mixture is different, the multimers formed by the molecules are structured in such a way that the average relative orientation of neighbouring dipoles is practically the same in both systems.

# 6. Conclusions

Measurements of $\rho$ ($T$ = 298.15 K), $c$ ($T$ = 298.15 K), $\varepsilon_r$ at 1 MHz ($T$ = 293.15 K to 303.15 K) and $n_D$ ($T$ = 293.15 K to 303.15 K) have been reported for the liquid mixtures methanol, propan-1-ol or pentan-1-ol + aniline at 0.1 MPa. The data have been useful to identify inconsistencies among different literature sources on alkan-1-ol + aniline systems. Positive $U_{m,V}^E$ and negative $\varepsilon_r^E$ that increase in absolute value with the length of the alkan-1-ol reveal the predominance of the breaking of the interactions between like molecules along mixing. This is supported by the application of the concentration-concentration structure factor formalism. The negative $V_m^E$ values show the existence of important structural effects in the mixtures under study. According to the Kirkwood-Fröhlich model, the average relative orientation of neighbouring dipoles is similar in the mixtures methanol + aniline or + pyridine.

## Funding

This work was supported by Consejería de Educación de Castilla y León, under Project VA100G19 (Apoyo a GIR, BDNS: 425389).

# References

[1] H. Matsuda, K. Ochi, K. Kojima, Determination and Correlation of LLE and SLE Data for the Methanol + Cyclohexane, Aniline + Heptane, and Phenol + Hexane System, J. Chem. Eng. Data 48 (2003) 184-189. https://doi.org/10.1021/je020156+

[2] P.J. Maher, B.D. Smith, Vapor-liquid equilibrium data for binary systems of aniline with acetone, acetonitrile, chlorobenzene, methanol, and 1-pentene, J. Chem. Eng. Data 25 (1980) 61-68. https://doi.org/10.1021/je60084a010

[3] E. Kristóf, I. Ország, F. Ratkovics, Static vapour-liquid equilibrium studies on {$xC_3H_7NH_2$ + (1−x)$C_3H_7OH$}, J. Chem. Thermodyn. 13 (1981) 557-562. https://doi.org/10.1016/0021-9614(81)90111-7

[4] J.A. González, I.G. de la Fuente, J.C. Cobos, Thermodynamics of mixtures with strongly negative deviations from Raoult's law. Part 3. Application of the DISQUAC model to mixtures of triethylamine with alkanols. Comparison with Dortmund UNIFAC and ERAS results, Can. J. Chem. 78 (2000) 1272-1284. https://doi.org/10.1139/v00-114

[5] J.A. González, I. García de la Fuente, J.C. Cobos, Thermodynamics of mixtures with strongly negative deviations from Raoult's Law: Part 4. Application of the DISQUAC model to mixtures of 1-alkanols with primary or secondary linear amines. Comparison with Dortmund UNIFAC and ERAS results, Fluid Phase Equilib. 168 (2000) 31-58. https://doi.org/10.1016/S0378-3812(99)00326-X

[6] S. Villa, N. Riesco, I. García de la Fuente, J.A. González, J.C. Cobos, Thermodynamics of mixtures with strongly negative deviations from Raoult's law: Part 5. Excess molar volumes at 298.15 K for 1-alkanols+dipropylamine systems: characterization in terms of the ERAS model, Fluid Phase Equilib. 190 (2001) 113-125. https://doi.org/10.1016/S0378-3812(01)00595-7

[7] S. Villa, N. Riesco, I. García de la fuente, J.A. González, J.C. Cobos, Thermodynamics of mixtures with strongly negative deviations from Raoult's law: Part 6. Excess molar volumes at 298.15 K for 1-alkanols + dibutylamine systems. Characterization in terms of the ERAS model, Fluid Phase Equilib. 198 (2002) 313-329. https://doi.org/10.1016/S0378-3812(01)00808-1

[8] S. Villa, N. Riesco, I.G. de la Fuente, J.A. González, J.C. Cobos, Thermodynamics of Mixtures with Strongly Negative Deviations from Raoult's Law. VII. Excess Molar Volumes at 25°C for 1-Alkanol + N-Methylbutylamine Systems—Characterization in Terms of the ERAS Model, J. Solution Chem. 32 (2003) 179-194. https://doi.org/10.1023/a:1022902323214

[9] S. Villa, N. Riesco, I. García de la Fuente, J.A. González, J.C. Cobos, Thermodynamics of mixtures with strongly negative deviations from Raoult's law. Part 8. Excess molar volumes at 298.15 K for 1-alkanol + isomeric amine ($C_6H_{15}N$) systems: Characterization in terms of the ERAS model, Fluid Phase Equilib. 216 (2004) 123-133. https://doi.org/10.1016/j.fluid.2003.10.008

[10] S. Villa, R. Garriga, P. Pérez, M. Gracia, J.A. González, I.G. de la Fuente, J.C. Cobos, Thermodynamics of mixtures with strongly negative deviations from Raoult's law: Part 9. Vapor–liquid equilibria for the system 1-propanol + di-n-propylamine at six temperatures between 293.15 and 318.15 K, Fluid Phase Equilib. 231 (2005) 211-220. https://doi.org/10.1016/j.fluid.2005.01.013

[11] L.F. Sanz, J.A. González, I. García De La Fuente, J.C. Cobos, Thermodynamics of mixtures with strongly negative deviations from Raoult's law. XI. Densities, viscosities and refractives indices at (293.15–303.15) K for cyclohexylamine + 1-propanol, or +1-butanol systems, J. Mol. Liq. 172 (2012) 26-33. https://doi.org/10.1016/j.molliq.2012.05.003

[12] L.F. Sanz, J.A. González, I. García de la Fuente, J.C. Cobos, Thermodynamics of mixtures with strongly negative deviations from Raoult's law. XII. Densities, viscosities and refractive indices at T = (293.15 to 303.15) K for (1-heptanol, or 1-decanol + cyclohexylamine) systems. Application of the ERAS model to (1-alkanol +

cyclohexylamine) mixtures, J. Chem. Thermodyn. 80 (2015) 161-171. https://doi.org/10.1016/j.jct.2014.09.005

[13] F. Hevia, J.A. González, A. Cobos, I. García de la Fuente, C. Alonso-Tristán, Thermodynamics of mixtures with strongly negative deviations from Raoult's law. XV. Permittivities and refractive indices for 1-alkanol + n-hexylamine systems at (293.15–303.15) K. Application of the Kirkwood-Fröhlich model, Fluid Phase Equilib. 468 (2018) 18-28. https://doi.org/10.1016/j.fluid.2018.04.007

[14] F. Hevia, A. Cobos, J.A. González, I. García de la Fuente, L.F. Sanz, Thermodynamics of mixtures with strongly negative deviations from Raoult's law. XVI. Permittivities and refractive indices for 1-alkanol + di-n-propylamine systems at (293.15–303.15) K. Application of the Kirkwood-Fröhlich model, J. Mol. Liq. 271 (2018) 704-714. https://doi.org/10.1016/j.molliq.2018.09.040

[15] F. Hevia, J.A. González, A. Cobos, I. García de la Fuente, L.F. Sanz, Thermodynamics of mixtures with strongly negative deviations from Raoult's law. XVII. Permittivities and refractive indices for alkan-1-ol + N,N-diethylethanamine systems at (293.15–303.15) K. Application of the Kirkwood-Fröhlich model, J. Chem. Thermodyn. 141 (2020) 105937. https://doi.org/10.1016/j.jct.2019.105937

[16] H. Fröhlich, Theory of Dielectrics, Clarendon Press, Oxford, 1958.

[17] A. Chelkowski, Dielectric Physics, Elsevier, Amsterdam, 1980.

[18] J.A. González, L.F. Sanz, I. García de la Fuente, J.C. Cobos, Thermodynamics of mixtures with strong negative deviations from Raoult's law. XIII. Relative permittivities for (1-alkanol + cyclohexylamine) systems, and dielectric study of (1-alkanol + polar) compound (amine, amide or ether) mixtures, J. Chem. Thermodyn. 91 (2015) 267-278. https://doi.org/10.1016/j.jct.2015.07.032

[19] L.F. Sanz, J.A. González, I.G. De La Fuente, J.C. Cobos, Thermodynamics of mixtures with strong negative deviations from raoult's law. XIV. density, permittivity, refractive index and viscosity data for the methanol + cyclohexylamine mixture at (293.15–303.15) K, Thermochim. Acta 631 (2016) 18-27. https://doi.org/10.1016/j.tca.2016.03.002

[20] D. Decroocq, Bull. Soc. Chim. Fr. (1964) 127. Cf. C. Wohlfahrt, Static Dielectric Constants of Pure Liquids and Binary Liquid Mixtures. Landolt-Börnstein - Group IV Physical Chemistry Vol. 117 (Supplement to IV/126). Springer Berlin Heidelberg, Berlin, 2008.

[21] K.M. Sumer, A.R. Thompson, Refraction, dispersion, and densities for methanol solutions of benzene, toluene, aniline, and phenol, J. Chem. Eng. Data 12 (1967) 489-493. https://doi.org/10.1021/je60035a007

[22] D.D. Deshpande, L.G. Bhatgadde, S. Oswal, C.S. Prabhu, Sound velocities and related properties in binary solutions of aniline, J. Chem. Eng. Data 16 (1971) 469-473. https://doi.org/10.1021/je60051a007

[23] M. Katz, P.W. Lobo, A.S. Miñano, H. Sólimo, Viscosities, Densities, and Refractive Indices of Binary Liquid Mixtures, Can. J. Chem. 49 (1971) 2605-2609. https://doi.org/10.1139/v71-429

[24] B.B. Swain, Acta Chim. Hung. 117 (1984) 383. Cf. C. Wohlfahrt, Static Dielectric Constants of Pure Liquids and Binary Liquid Mixtures. Landolt-Börnstein - Group IV Physical Chemistry Vol. 317 (Supplement to IV/386). Springer Berlin Heidelberg, Berlin, 2008.

[25] S.K. Ray, G.S. Roy, Proc. Natl. Sci. Acad. India A 59 (1993) 205. Cf. C. Wohlfahrt, Static Dielectric Constants of Pure Liquids and Binary Liquid Mixtures. Landolt-Börnstein - Group IV Physical Chemistry Vol. 217 (Supplement to IV/206). Springer Berlin Heidelberg, Berlin, 2008.

[26] R.H. Fattepur, M.T. Hosamani, D.K. Deshpande, S.C. Mehrotra, Dielectric relaxation and structural study of aniline–methanol mixture using picosecond time domain reflectometry, J. Chem. Phys. 101 (1994) 9956-9960. https://doi.org/10.1063/1.467897

[27] K. Garabadu, B.B. Swain, Indian J. Phys. B 68 (1994) 271. Cf. C. Wohlfahrt, Static Dielectric Constants of Pure Liquids and Binary Liquid Mixtures. Landolt-Börnstein -

Group IV Physical Chemistry Vol. 217 (Supplement to IV/276). Springer Berlin Heidelberg, Berlin, 2008.

[28] S.P. Patil, A.S. Chaudhari, M.P. Lokhande, M.K. Lande, A.G. Shankarwar, S.N. Helambe, B.R. Arbad, S.C. Mehrotra, Dielectric Measurements of Aniline and Alcohol Mixtures at 283, 293, 303, and 313 K Using the Time Domain Technique, J. Chem. Eng. Data 44 (1999) 875-878. https://doi.org/10.1021/je980250j

[29] S.K. Ray, J. Rath, S.K. Chakrabarty, C. Dwivedi, Acta Cienc. Indica P 26 (2000) 33. Cf. C. Wohlfahrt, Static Dielectric Constants of Pure Liquids and Binary Liquid Mixtures. Landolt-Börnstein - Group IV Physical Chemistry Vol. 17 (Supplement to IV/36). Springer Berlin Heidelberg, Berlin, 2008.

[30] N.G. Tsierkezos, M.M. Palaiologou, I.E. Molinou, Densities and Viscosities of 1-Pentanol Binary Mixtures at 293.15 K, J. Chem. Eng. Data 45 (2000) 272-275. https://doi.org/10.1021/je9902138

[31] M.A. Saleh, M. Alauddin, S. Begum, Excess Molar Volume of 1-propanol+Aniline, +N-methylaniline, +N,N-dimethylaniline, Phys. Chem. Liq. 39 (2001) 453-464. https://doi.org/10.1080/00319100108031676

[32] V.A. Rana, A.D. Vyas, S.C. Mehrotra, Dielectric relaxation study of mixtures of 1-propanol with aniline, 2-chloroaniline and 3-chloroaniline at different temperatures using time domain reflectometry, J. Mol. Liq. 102 (2003) 379-391. https://doi.org/10.1016/S0167-7322(02)00162-9

[33] M.A. Wahab, M.A. Islam, M.A. Ali, M.A. Mottaleb, Evaluation of Excess Molar Volumes of n-butanol with nitrobenzene-aniline-acetonitrile Binary Liquid Mixtures at Temperatures of 298.15, 303.15, 308.15, and 313.15 K, Phys. Chem. Liq. 41 (2003) 189-195. https://doi.org/10.1080/00319100307953

[34] A. Goyal, M. Singh, Densities, viscosities and thermodynamic excess properties of ternary liquid mixtures of aniline and methyl alcohol as common components and non-polar solvents at 298.15 K, Ind. J. Chem. Section a 46 (2007) 60-69.

[35] A.K. Nain, Densities and Volumetric Properties of Binary Mixtures of Aniline with 1-Propanol, 2-Propanol, 2-Methyl-1-Propanol, and 2-Methyl-2-Propanol at Temperatures from 293.15 to 318.15 K, Int. J. Thermophys. 28 (2007) 1228-1244. https://doi.org/10.1007/s10765-007-0204-0

[36] A.K. Nain, Ultrasonic and viscometric study of molecular interactions in binary mixtures of aniline with 1-propanol, 2-propanol, 2-methyl-1-propanol, and 2-methyl-2-propanol at different temperatures, Fluid Phase Equilib. 259 (2007) 218-227. https://doi.org/10.1016/j.fluid.2007.07.016

[37] A.K. Nain, Deviations in refractive indices and applicability of mixing rules in aniline + alkanol binary mixtures at different temperatures, Phys. Chem. Liq. 48 (2010) 41-49. https://doi.org/10.1080/00319100802641815

[38] M. Swetha Sandhya, P. Biswas, N.R. Vinay, K. Sivakumar, R. Dey, Molecular interaction studies based on transport, thermodynamic and excess properties of aniline and alkanol mixtures at varying temperatures, J. Mol. Liq. 278 (2019) 219-225. https://doi.org/10.1016/j.molliq.2019.01.056

[39] CIAAW, Atomic weights of the elements 2015, ciaaw.org/atomic-weights.htm (accessed 2015).

[40] J.A. González, I. Alonso, I. Mozo, I. García de la Fuente, J.C. Cobos, Thermodynamics of (ketone + amine) mixtures. Part VI. Volumetric and speed of sound data at (293.15, 298.15, and 303.15) K for (2-heptanone + dipropylamine, +dibutylamine, or +triethylamine) systems, J. Chem. Thermodyn. 43 (2011) 1506-1514. https://doi.org/10.1016/j.jct.2011.05.003

[41] D. Schneditz, T. Kenner, H. Heimel, H. Stabinger, A sound-speed sensor for the measurement of total protein concentration in disposable, blood-perfused tubes, J. Acoust. Soc. Am. 86 (1989) 2073-2080. https://doi.org/10.1121/1.398466

[42] E. Junquera, G. Tardajos, E. Aicart, Speeds of sound and isentropic compressibilities of (cyclohexane + benzene and (1-chlorobutane + n-hexane or n-heptane or n-octane or n-

decane) at 298.15 K, J. Chem. Thermodyn. 20 (1988) 1461-1467. https://doi.org/10.1016/0021-9614(88)90041-9

[43] K. Tamura, K. Ohomuro, S. Murakami, Speeds of sound, isentropic and isothermal compressibilities, and isochoric heat capacities of {xc-C6H12+(1−x)C6H6}, x{CCl4+(1−x)C6H6}, and x{C7H16+(1−x)C6H6} at 298.15 K, J. Chem. Thermodyn. 15 (1983) 859-868. https://doi.org/10.1016/0021-9614(83)90092-7

[44] K. Tamura, S. Murakami, Speeds of sound, isentropic and isothermal compressibilities, and isochoric heat capacities of {xc-C6H12 + (1 − x)C6H6} from 293.15 to 303.15 K, J. Chem. Thermodyn. 16 (1984) 33-38. https://doi.org/10.1016/0021-9614(84)90072-7

[45] Y.P. Handa, G.C. Benson, Volume changes on mixing two liquids: A review of the experimental techniques and the literature data, Fluid Phase Equilib. 3 (1979) 185-249. https://doi.org/10.1016/0378-3812(79)85010-4

[46] K.N. Marsh, Recommended reference materials for the realization of physicochemical properties, Blackwell Scientific Publications, Oxford, UK, 1987.

[47] V. Alonso, J.A. González, I. García de la Fuente, J.C. Cobos, Dielectric and refractive index measurements for the systems 1-pentanol + octane, or + dibutyl ether or for dibutyl ether + octane at different temperatures, Thermochim. Acta 543 (2012) 246-253. https://doi.org/10.1016/j.tca.2012.05.036

[48] G.C. Benson, C.J. Halpin, A.J. Treszczanowicz, Excess volumes and isentropic compressibilities for (2-ethoxyethanol + n-heptane) at 298.15 K, J. Chem. Thermodyn. 13 (1981) 1175-1183. https://doi.org/10.1016/0021-9614(81)90017-3

[49] G. Douhéret, M.I. Davis, J.C.R. Reis, M.J. Blandamer, Isentropic Compressibilities—Experimental Origin and the Quest for their Rigorous Estimation in Thermodynamically Ideal Liquid Mixtures, ChemPhysChem 2 (2001) 148-161. https://doi.org/10.1002/1439-7641(20010316)2:3<148::AID-CPHC148>3.0.CO;2-J

[50] G. Douhéret, C. Moreau, A. Viallard, Excess thermodynamic quantities in binary systems of non electrolytes.: Different ways of calculating excess compressibilities, Fluid Phase Equilib. 22 (1985) 277-287. https://doi.org/10.1016/0378-3812(85)87027-8

[51] J.C.R. Reis, T.P. Iglesias, G. Douhéret, M.I. Davis, The permittivity of thermodynamically ideal liquid mixtures and the excess relative permittivity of binary dielectrics, Phys. Chem. Chem. Phys. 11 (2009) 3977-3986. https://doi.org/10.1039/B820613A

[52] J.C.R. Reis, I.M.S. Lampreia, Â.F.S. Santos, M.L.C.J. Moita, G. Douhéret, Refractive Index of Liquid Mixtures: Theory and Experiment, ChemPhysChem 11 (2010) 3722-3733. https://doi.org/10.1002/cphc.201000566

[53] O. Redlich, A.T. Kister, Algebraic Representation of Thermodynamic Properties and the Classification of Solutions, Ind. & Eng. Chem. 40 (1948) 345-348. https://doi.org/10.1021/ie50458a036

[54] P.R. Bevington, D.K. Robinson, Data Reduction and Error Analysis for the Physical Sciences, McGraw-Hill, New York, 2000.

[55] Y.X. Wang, J.P. Chao, M. Dai, Studies on the thermodynamic properties of binary mixtures containing an alcohol: XIII. Excess molar enthalpies of some mixtures of n-alcohols with aniline or chlorobenzene at 298.15 K, Thermochim. Acta 169 (1990) 161-169. https://doi.org/10.1016/0040-6031(90)80142-L

[56] J.P. Chao, M. Dai, Studies on thermodynamic properties of binary mixtures containing an alcohol XI. Excess enthalpies of each of (one of the four butanols + chlorobenzene or aniline), J. Chem. Thermodyn. 21 (1989) 337-342. https://doi.org/10.1016/0021-9614(89)90134-1

[57] J.S. Rowlinson, F.L. Swinton, Liquids and Liquid Mixtures, 3rd Edition, Butterworths, G. B., 1982.

[58] J.A. Riddick, W.B. Bunger, T.K. Sakano, Organic solvents: physical properties and methods of purification, Wiley, New York, 1986.

[59] A. Heintz, P.K. Naicker, S.P. Verevkin, R. Pfestorf, Thermodynamics of alkanol + amine mixtures. Experimental results and ERAS model calculations of the heat of

mixing, Ber. Bunsenges. Phys. Chem. 102 (1998) 953-959. https://doi.org/10.1002/bbpc.19981020707

[60] J.A. González, I. Mozo, I. García de la Fuente, J.C. Cobos, Thermodynamics of organic mixtures containing amines. IV. Systems with aniline, Can. J. Chem. 83 (2005) 1812-1825. https://doi.org/10.1139/v05-190

[61] I. Alonso, V. Alonso, I. Mozo, I. García de la Fuente, J.A. González, J.C. Cobos, Thermodynamics of Ketone + Amine Mixtures. I. Volumetric and Speed of Sound Data at (293.15, 298.15, and 303.15) K for 2-Propanone + Aniline, + N-Methylaniline, or + Pyridine Systems, J. Chem. Eng. Data 55 (2010) 2505-2511. https://doi.org/10.1021/je900874z

[62] L. Venkatramana, R.L. Gardas, K. Sivakumar, K. Dayananda Reddy, Thermodynamics of binary mixtures: The effect of substituents in aromatics on their excess properties with benzylalcohol, Fluid Phase Equilib. 367 (2014) 7-21. https://doi.org/10.1016/j.fluid.2014.01.019

[63] M.T. Rätzsch, C. Wohlfarth, M. Claudius, J. Prakt. Chemie 319 (1977) 353.

[64] N.V. Sastry, M.K. Valand, Densities, Speeds of Sound, Viscosities, and Relative Permittivities for 1-Propanol + and 1-Butanol + Heptane at 298.15 K and 308.15 K, J. Chem. Eng. Data 41 (1996) 1421-1425. https://doi.org/10.1021/je960135d

[65] N.V. Sastry, M.K. Valand, Dielectric constants, refractive indexes and polarizations for 1-Alcohol +Heptane mixtures at 298.15 and 308.15 K, Ber. Bunsenges. Phys. Chem. 101 (1997) 243-250. https://doi.org/10.1002/bbpc.19971010212

[66] A. Chaudhari, S.C. Mehrotra, Dielectric relaxation study of pyridine-alcohol mixtures using time domain reflectometry, Mol. Phys. 100 (2002) 3907-3913. https://doi.org/10.1080/0026897021000023668

[67] O. Dahmani, A. Ait-Kaci, Diagramme d'equilibre liquide-liquide de systemes binaires pyridine+n-alcanes, J. Thermal Anal. 44 (1995) 385-393. https://doi.org/10.1007/BF02636129

[68] M.S. Bakshi, G. Kaur, Thermodynamic Behavior of Mixtures. 4. Mixtures of Methanol with Pyridine and N,N-Dimethylformamide at 25 °C, J. Chem. Eng. Data 42 (1997) 298-300. https://doi.org/10.1021/je960300p

[69] H. Touhara, K. Nakanishi, Excess molar enthalpies of methanol + pyridine, + methylpyridine, and + 2,6-dimethylpyridine, J. Chem. Thermodyn. 17 (1985) 909-914. https://doi.org/10.1016/0021-9614(85)90002-3

[70] J.C. Cobos, An exact quasi-chemical equation for excess heat capacity with W-shaped concentration dependence, Fluid Phase Equilib. 133 (1997) 105-127. https://doi.org/10.1016/S0378-3812(97)00012-5

[71] R.G. Rubio, M. Cáceres, R.M. Masegosa, L. Andreolli-Ball, M. Costas, D. Patterson, Mixtures with "w-Shape" CEp curves. A light scattering study, Ber. Bunsenges. Phys. Chem. 93 (1989) 48-56. https://doi.org/10.1002/bbpc.19890930110

[72] J.A. González, J.C. Cobos, I. García de la Fuente, I. Mozo, Thermodynamics of mixtures containing amines. IX. Application of the concentration–concentration structure factor to the study of binary mixtures containing pyridines, Thermochim. Acta 494 (2009) 54-64. https://doi.org/10.1016/j.tca.2009.04.017

[73] P. Brocos, A. Piñeiro, R. Bravo, A. Amigo, Refractive indices, molar volumes and molar refractions of binary liquid mixtures: concepts and correlations, Phys. Chem. Chem. Phys. 5 (2003) 550-557. https://doi.org/10.1039/B208765K

[74] J.C.R. Reis, T.P. Iglesias, Kirkwood correlation factors in liquid mixtures from an extended Onsager-Kirkwood-Frohlich equation, Phys. Chem. Chem. Phys. 13 (2011) 10670-10680. https://doi.org/10.1039/C1CP20142E

[75] C. Moreau, G. Douhéret, Thermodynamic and physical behaviour of water + acetonitrile mixtures. Dielectric properties, J. Chem. Thermodyn. 8 (1976) 403-410. https://doi.org/10.1016/0021-9614(76)90060-4

[76] Y. Marcus, The structuredness of solvents, J. Solution Chem. 21 (1992) 1217-1230. https://doi.org/10.1007/bf00667218

[77] J.A. González, C.A. Tristán, F. Hevia, I.G. De La Fuente, L.F. Sanz, Thermodynamics of mixtures containing aromatic nitriles, J. Chem. Thermodyn. 116 (2018) 259-272. https://doi.org/10.1016/j.jct.2017.09.027

[78] F. Hevia, J.A. González, A. Cobos, I. García de la Fuente, L.F. Sanz, Thermodynamics of amide + amine mixtures. 4. Relative permittivities of N,N-dimethylacetamide + N-propylpropan-1-amine, + N-butylbutan-1-amine, + butan-1-amine, or + hexan-1-amine systems and of N,N-dimethylformamide + aniline mixture at several temperatures. Characterization of amine + amide systems using ERAS, J. Chem. Thermodyn. 118 (2018) 175-187. https://doi.org/10.1016/j.jct.2017.11.011

[79] M. El-Hefnawy, K. Sameshima, T. Matsushita, R. Tanaka, Apparent Dipole Moments of 1-Alkanols in Cyclohexane and n-Heptane, and Excess Molar Volumes of (1-Alkanol + Cyclohexane or n-Heptane) at 298.15 K, J. Solution Chem. 34 (2005) 43-69. https://doi.org/10.1007/s10953-005-2072-1

[80] G.J. Janz, R.P.T. Tomkins, Nonaqueus Electrolytes Handbook, Vol. 1, Academic Press, New York, 1972.

[81] S.P. Serbanovic, M.L. Kijevcanin, I.R. Radovic, B.D. Djordjevic, Effect of temperature on the excess molar volumes of some alcohol + aromatic mixtures and modelling by cubic EOS mixing rules, Fluid Phase Equilib. 239 (2006) 69-82. https://doi.org/10.1016/j.fluid.2005.10.022

[82] J.L. Hales, J.H. Ellender, Liquid densities from 293 to 490 K of nine aliphatic alcohols, J. Chem. Thermodyn. 8 (1976) 1177-1184. https://doi.org/10.1016/0021-9614(76)90126-9

[83] G.A. Iglesias-Silva, A. Guzmán-López, G. Pérez-Durán, M. Ramos-Estrada, Densities and Viscosities for Binary Liquid Mixtures of n-Undecane + 1-Propanol, + 1-Butanol, + 1-Pentanol, and + 1-Hexanol from 283.15 to 363.15 K at 0.1 MPa, J. Chem. Eng. Data 61 (2016) 2682-2699. https://doi.org/10.1021/acs.jced.6b00121

[84] D. Soldatović, J. Vuksanović, I. Radović, Z. Višak, M. Kijevčanin, Excess molar volumes and viscosity behaviour of binary mixtures of aniline/or N,N-dimethylaniline with imidazolium ionic liquids having triflate or bistriflamide anion, J. Chem. Thermodyn. 109 (2017) 137-154. https://doi.org/10.1016/j.jct.2017.02.007

[85] M.I. Aralaguppi, C.V. Jadar, T.M. Aminabhavi, Density, Viscosity, Refractive Index, and Speed of Sound in Binary Mixtures of Acrylonitrile with Methanol, Ethanol, Propan-1-ol, Butan-1-ol, Pentan-1-ol, Hexan-1-ol, Heptan-1-ol, and Butan-2-ol, J. Chem. Eng. Data 44 (1999) 216-221. https://doi.org/10.1021/je9802219

[86] R. Anwar Naushad, S. Yasmeen, Volumetric, compressibility and viscosity studies of binary mixtures of [EMIM][NTf2] with ethylacetate/methanol at (298.15–323.15) K, J. Mol. Liq. 224, Part A (2016) 189-200. https://doi.org/10.1016/j.molliq.2016.09.077

[87] A. Rodríguez, J. Canosa, J. Tojo, Density, Refractive Index, and Speed of Sound of Binary Mixtures (Diethyl Carbonate + Alcohols) at Several Temperatures, J. Chem. Eng. Data 46 (2001) 1506-1515. https://doi.org/10.1021/je010148d

[88] B. González, Á. Domínguez, J. Tojo, Viscosity, density, and speed of sound of methylcyclopentane with primary and secondary alcohols at T=(293.15, 298.15, and 303.15)K, J. Chem. Thermodyn. 38 (2006) 1172-1185. https://doi.org/10.1016/j.jct.2005.12.010

[89] J.M. Resa, C. González, S. Ortiz de Landaluce, J. Lanz, (Vapour + liquid) equilibria, densities, excess molar volumes, refractive indices, speed of sound for (methanol + allyl acetate) and (vinyl acetate+ allyl acetate), J. Chem. Thermodyn. 34 (2002) 1013-1027. https://doi.org/10.1006/jcht.2002.0979

[90] M.T. Zafarani-Moattar, H. Shekaari, Volumetric and Speed of Sound of Ionic Liquid, 1-Butyl-3-methylimidazolium Hexafluorophosphate with Acetonitrile and Methanol at T = (298.15 to 318.15) K, J. Chem. Eng. Data 50 (2005) 1694-1699. https://doi.org/10.1021/je050165t

[91] M.A. Villamañan, C. Casanova, G. Roux-Desgranges, J.-P.E. Grolier, Thermochemical behaviour of mixtures of n-alcohol + aliphatic ether: heat capacities and volumes at

298.15 K, Thermochim. Acta 52 (1982) 279-283. https://doi.org/10.1016/0040-6031(82)85206-4

[92] G.C. Benson, P.J. D'Arcy, Excess isobaric heat capacities of some binary mixtures: (a C5-alkanol + n-heptane) at 298.15 K, J. Chem. Thermodyn. 18 (1986) 493-498. https://doi.org/10.1016/0021-9614(86)90099-6

[93] N. Nichols, I. Wadsö, Thermochemistry of solutions of biochemical model compounds 3. Some benzene derivatives in aqueous solution, 7 (1975) 329-336. https://doi.org/10.1016/0021-9614(75)90169-X

[94] J. Canosa, A. Rodríguez, J. Tojo, Binary mixture properties of diethyl ether with alcohols and alkanes from 288.15 K to 298.15 K, Fluid Phase Equilib. 156 (1999) 57-71. https://doi.org/10.1016/S0378-3812(99)00032-1

[95] M.J. Fontao, M. Iglesias, Effect of Temperature on the Refractive Index of Aliphatic Hydroxilic Mixtures (C2–C3), Int. J. Thermophys. 23 (2002) 513-527. https://doi.org/10.1023/A:1015113604024

[96] R.S. Ramadevi, M.V.P. Rao, Excess Volumes of Substituted Benzenes with N,N-Dimethylformamide, J. Chem. Eng. Data 40 (1995) 65-67. https://doi.org/10.1021/je00017a013

[97] S. Chen, Q. Lei, W. Fang, Density and Refractive Index at 298.15 K and Vapor−Liquid Equilibria at 101.3 kPa for Four Binary Systems of Methanol, n-Propanol, n-Butanol, or Isobutanol with N-Methylpiperazine, J. Chem. Eng. Data 47 (2002) 811-815. https://doi.org/10.1021/je010249b

[98] S. Kumar, A. Maken, S. Agarwal, S. Maken, Topological studies of molecular interactions of 1,4-dioxane with formamides or anilines at 308.15K, J. Mol. Liq. 155 (2010) 115-120. https://doi.org/10.1016/j.molliq.2010.05.023

[99] S.M. Pereira, T.P. Iglesias, J.L. Legido, L. Rodríguez, J. Vijande, Changes of refractive index on mixing for the binary mixtures {xCH3OH+(1−x)CH3OCH2(CH2OCH2)3CH2OCH3} and {xCH3OH+(1−x)CH3OCH2(CH2OCH2)nCH2OCH3} (n=3−9) at temperatures from 293.15 K to 333.15 K, J. Chem. Thermodyn. 30 (1998) 1279-1287. https://doi.org/10.1006/jcht.1998.0395

[100] M.N.M. Al-Hayan, Densities, excess molar volumes, and refractive indices of 1,1,2,2-tetrachloroethane and 1-alkanols binary mixtures, J. Chem. Thermodyn. 38 (2006) 427-433. https://doi.org/10.1016/j.jct.2005.06.015

[101] R.D. Bezman, E.F. Casassa, R.L. Kay, The temperature dependence of the dielectric constants of alkanols, 73–74 (1997) 397-402. https://doi.org/10.1016/S0167-7322(97)00082-2

[102] A.P. Gregory, R.N. Clarke, Traceable measurements of the static permittivity of dielectric reference liquids over the temperature range 5–50 °C, Meas. Sci. Technol. 16 (2005) 1506-1516. https://doi.org/10.1088/0957-0233/16/7/013

[103] E.G. Cowley, 680. The dielectric polarisation of solutions. Part I. The polarisations and apparent dipole moments of various primary, secondary, and tertiary amines in solutions in non-polar solvents and in the liquid state, J. Chem. Soc. (Resumed) (1952) 3557-3570. https://doi.org/10.1039/JR9520003557

[104] E. Fischer, Dielektrische Relaxationsuntersuchungen zur Frage des inneren Feldes und zur Charakterisierung von assoziierten und nicht assoziierten Dipolflüssigkeiten, Z. Phys. 127 (1950) 49-71. https://doi.org/10.1007/BF01338983

[105] T.P. Iglesias, J.L. Legido, S.M. Pereira, B. de Cominges, M.I. Paz Andrade, Relative permittivities and refractive indices on mixing for (n-hexane + 1-pentanol, or 1-hexanol, or 1-heptanol ) at T = 298.15 K, J. Chem. Thermodyn. 32 (2000) 923-930. https://doi.org/10.1006/jcht.2000.0661

[106] S. Lata, K.C. Singh, A. Suman, A study of dielectric properties and refractive indices of aniline + benzene, + toluene, + o-xylene, and + p-xylene at 298.15 K, J. Mol. Liq. 147 (2009) 191-197. https://doi.org/10.1016/j.molliq.2009.04.002

[107] A.N. Prajapati, V.A. Rana, A.D. Vyas, Study of molecular interaction between some primary alcohols and anilines using concentration dependent dielectric properties, Ind. J. Pure Appl. Phys. 51 (2013) 104-111.

[108] G. Parthipan, T. Thenappan, Dielectric and thermodynamic behavior of binary mixture of anisole with morpholine and aniline at different temperatures, J. Mol. Liq. 138 (2008) 20-25. https://doi.org/10.1016/j.molliq.2007.06.010

Table 1

Sample description.

| Chemical name | CAS Number | Source | Purification method | Purity[a] | Water content[b] |
|---|---|---|---|---|---|
| methanol | 67-56-1 | Sigma-Aldrich | none | 0.999 | $2 \cdot 10^{-5}$ |
| propan-1-ol | 71-23-8 | Fluka | none | 0.999 | $1 \cdot 10^{-3}$ |
| pentan-1-ol | 71-41-0 | Sigma-Aldrich | none | 0.999 | $3 \cdot 10^{-4}$ |
| aniline | 62-53-3 | Sigma-Aldrich | none | 0.999 | $4 \cdot 10^{-4}$ |

[a] In mole fraction. By gas chromatography. Provided by the supplier.

[b] In mass fraction. By Karl-Fischer titration.

Table 2

Thermophysical properties of the pure liquids used in this work at temperature $T$ and pressure $p$ = 0.1 MPa: dipole moment ($\mu$), density ($\rho^*$), speed of sound ($c^*$), isobaric thermal expansion coefficient ($\alpha_p^*$), isentropic compressibility ($\kappa_S^*$), molar isobaric heat capacity ($C_{pm}^*$), isothermal compressibility ($\kappa_T^*$), refractive index at the sodium D-line ($n_D^*$) and relative permittivity at frequency $f$ = 1 MHz ($\varepsilon_r^*$). [a]

| Property | $T$/K | Methanol | propan-1-ol | pentan-1-ol | aniline |
|---|---|---|---|---|---|
| $\mu$/D | | 1.664 [79] | 1.629 [79] | 1.598 [79] | 1.51 [58] |
| $\rho^*$/g·cm$^{-3}$ | 293.15 | 0.79191 | 0.80352 | 0.81454 | 1.02166 |
| | | 0.7916 [80] | 0.80361 [82] | 0.81468 [83] | 1.02166 [30] |
| | | 0.791400 [81] | | | 1.0217 [84] |
| | 298.15 | 0.78682 | 0.79941 | 0.81081 | 1.01731 |
| | | 0.7869 [85] | 0.79960 [82] | 0.81103 [83] | 1.0174 [84] |
| | | 0.786884 [86] | | | |
| | 303.15 | 0.78259 | 0.79547 | 0.80724 | 1.01301 |
| | | 0.782158 [86] | 0.79561 [82] | 0.81737 [83] | 1.0130 [84] |
| $c^*$/m·s$^{-1}$ | 293.15 | 1119.1 | 1222.5 | 1292.4 | 1658.0 |
| | | 1119 [87] | 1223 [88] | 1292 [87] | 1651.3 [36] |
| | | | | | 1657.0 [61] |
| | | | | | 1657.0 [22] |
| | 298.15 | 1102.1 | 1205.1 | 1275.3 | 1638.6 |
| | | 1101.9 [89] | 1206 [88] | 1276 [87] | 1632.8 [36] |
| | | | | | 1638.6 [61] |
| | 303.15 | 1086.6 | 1188.6 | 1259.0 | 1617.9 |
| | | 1186.37 [90] | 1189 [88] | 1259 [87] | 1614.5 [36] |
| | | | | | 1619.2 [61] |
| $\alpha_p^*$/10$^{-3}$K$^{-1}$ | 298.15 | 1.185 | 1.007 | 0.900 | 0.850 |
| | | 1.196 [58] | 1.004 [58] | 0.905 [58] | 0.849 [58] |
| $\kappa_S^*$/TPa$^{-1}$ | 293.15 | 1008.2 | 832.7 | 735.0 | 356.1 |
| | 298.15 | 1046.7 | 861.4 | 758.3 | 366.1 |
| | | 1028 [58] | 849 [58] | | 368 [58] |
| | 303.15 | 1082.3 | 889.8 | 781.5 | 377.1 |
| $C_{pm}^*$/J·mol$^{-1}$·K$^{-1}$ | 298.15 | 81.92 [91] | 146.88 [91] | 207.45 [92] | 191.01 [93] |
| [b] $\kappa_T^*$/TPa$^{-1}$ | 298.15 | 1254.9 | 1016.1 | 884.9 | 469.3 |
| | | 1248 [58] | 1026 [58] | 884 [58] | |
| $n_D^*$ | 293.15 | 1.32852 | 1.38513 | 1.41000 | 1.58656 |
| | | 1.32859 [94] | 1.38512 [95] | 1.40986 [87] | 1.5865 [96] |
| | | | | | 1.58660 [21] |
| | 298.15 | 1.32639 | 1.38309 | 1.40800 | 1.58378 |
| | | 1.32652 [97] | 1.38307 [87] | 1.40789 [87] | 1.58364 [58] |
| | | | | | 1.5836 [98] |
| | 303.15 | 1.32431 | 1.38101 | 1.40596 | 1.58109 |
| | | 1.32457 [99] | 1.38104 [87] | 1.40592 [100] | 1.58143 [21] |
| | | 1.32410 [87] | | | |
| $\varepsilon_r^*$ | 293.15 | 33.576 | 21.225 | 15.746 | 7.137 |
| | | 33.61 [101] | 21.15 [102] | 15.63 [101] | 7.07 [103] |
| | | | | | 7.045 [104] |

| | | | | |
|---|---|---|---|---|
| 298.15 | 32.624 | 20.545 | 15.162 | 7.004 |
| | 32.62 [101] | 20.42 [102] | 15.08 [105] | 6.940 [104] |
| | | | | 6.774 [106] |
| 303.15 | 31.684 | 19.873 | 14.586 | 6.876 |
| | 31.66 [101] | 19.75 [102] | 14.44 [101] | 6.88 [107] |
| | | | | 6.857 [108] |

[a] Standard uncertainties ($u$): $u(T)$ = 0.01 K for $\rho^*$ measurements; $u(T)$ = 0.02 K for $\varepsilon_r^*$ and $n_D^*$ measurements; $u(p)$ = 1 kPa; $u(f)$ = 20 Hz; $u(c^*)$ = 0.2 m·s$^{-1}$; $u(n_D^*)$ = 0.00008. Relative standard uncertainties ($u_r$): $u_r(\rho^*)$ = 0.0008; $u_r(\alpha_p^*)$ = 0.028; $u_r(\kappa_S^*)$ = 0.002; $u_r(\kappa_T^*)$ = 0.015; $u_r(\varepsilon_r^*)$ = 0.003.

[b] Determined using experimental values measured in this work and $C_{pm}^*$ values from the literature included in this table.

Table 3

Density ($\rho$), speed of sound ($c$), excess molar volume ($V_m^E$), excess isentropic compressibility ($\kappa_S^E$) and excess speed of sound ($c^E$) of alkan-1-ol (1) + aniline (2) liquid mixtures as functions of the mole fraction of the alkan-1-ol ($x_1$) at temperature $T$ = 298.15 K and pressure $p$ = 0.1 MPa.[a]

| $x_1$ | $\rho$ /g·cm$^{-3}$ | $c$ /m·s$^{-1}$ | $V_m^E$ /cm$^3$·mol$^{-1}$ | $\kappa_S^E$ /TPa$^{-1}$ | $c^E$ /m·s$^{-1}$ | $x_1$ | $\rho$ /g·cm$^{-3}$ | $c$ /m·s$^{-1}$ | $V_m^E$ /cm$^3$·mol$^{-1}$ | $\kappa_S^E$ /TPa$^{-1}$ | $c^E$ /m·s$^{-1}$ |
|---|---|---|---|---|---|---|---|---|---|---|---|
| \multicolumn{12}{c}{methanol (1) + aniline (2) ; $T$/K = 298.15} |
| 0.0502 | 1.01351 | 1629.5 | −0.1313 | −10.6 | 21.5 | 0.5508 | 0.94981 | 1464.9 | −0.9268 | −120.3 | 142.5 |
| 0.1006 | 1.00939 | 1618.6 | −0.2573 | −21.1 | 41.0 | 0.5977 | 0.94018 | 1441.1 | −0.9491 | −129.6 | 143.6 |
| 0.1516 | 1.00483 | 1606.5 | −0.3749 | −31.9 | 59.1 | 0.6507 | 0.92770 | 1410.9 | −0.9354 | −138.0 | 141.2 |
| 0.2003 | 1.00006 | 1593.9 | −0.4765 | −42.5 | 75.3 | 0.6971 | 0.91554 | 1381.9 | −0.9090 | −143.5 | 136.2 |
| 0.2510 | 0.99463 | 1579.6 | −0.5722 | −53.7 | 90.3 | 0.7501 | 0.90000 | 1345.8 | −0.8596 | −146.8 | 127.0 |
| 0.3024 | 0.98872 | 1564.0 | −0.6681 | −65.4 | 104.2 | 0.7997 | 0.88342 | 1307.8 | −0.7821 | −144.4 | 113.7 |
| 0.3489 | 0.98273 | 1548.5 | −0.7352 | −75.8 | 115.0 | 0.8497 | 0.86434 | 1265.3 | −0.6684 | −134.6 | 95.5 |
| 0.3976 | 0.97597 | 1530.9 | −0.8022 | −86.9 | 124.7 | 0.8992 | 0.84266 | 1218.1 | −0.5141 | −113.2 | 71.6 |
| 0.4504 | 0.96783 | 1510.1 | −0.8582 | −98.8 | 133.2 | 0.9500 | 0.81677 | 1163.5 | −0.2948 | −72.3 | 40.0 |
| 0.4976 | 0.95983 | 1489.8 | −0.8984 | −109.1 | 138.6 | | | | | | |
| \multicolumn{12}{c}{propan-1-ol (1) + aniline (2) ; $T$/K = 298.15} |
| 0.0498 | 1.00931 | 1621.9 | −0.0891 | −10.1 | 20.6 | 0.5494 | 0.91498 | 1421.7 | −0.6034 | −74.9 | 84.4 |
| 0.0994 | 1.00114 | 1604.0 | −0.1741 | −19.5 | 37.3 | 0.5998 | 0.90391 | 1399.9 | −0.6167 | −76.5 | 81.2 |
| 0.1508 | 0.99242 | 1585.0 | −0.2551 | −28.8 | 51.8 | 0.6507 | 0.89215 | 1377.0 | −0.5990 | −76.1 | 76.1 |
| 0.2002 | 0.98379 | 1566.6 | −0.3260 | −37.3 | 63.3 | 0.7005 | 0.88038 | 1354.5 | −0.5789 | −74.3 | 69.9 |
| 0.2478 | 0.97524 | 1547.6 | −0.3882 | −44.5 | 71.4 | 0.7505 | 0.86807 | 1331.4 | −0.5358 | −70.2 | 62.3 |
| 0.3001 | 0.96555 | 1526.7 | −0.4472 | −52.0 | 78.5 | 0.8004 | 0.85548 | 1308.1 | −0.4872 | −64.1 | 53.5 |
| 0.3518 | 0.95566 | 1505.6 | −0.4961 | −58.6 | 83.3 | 0.8498 | 0.84258 | 1284.5 | −0.4218 | −55.4 | 43.4 |
| 0.4015 | 0.94592 | 1485.0 | −0.5399 | −64.2 | 86.0 | 0.8998 | 0.82877 | 1259.3 | −0.3096 | −42.1 | 31.0 |
| 0.4500 | 0.93605 | 1464.5 | −0.5673 | −68.7 | 87.0 | 0.9502 | 0.81434 | 1232.8 | −0.1733 | −24.1 | 16.6 |
| 0.5003 | 0.92564 | 1443.2 | −0.5987 | −72.6 | 86.6 | | | | | | |
| \multicolumn{12}{c}{pentan-1-ol (1) + aniline (2) ; $T$/K = 298.15} |
| 0.0574 | 1.00395 | 1611.6 | −0.0522 | −9.1 | 18.4 | 0.5470 | 0.89781 | | −0.2430 | | |
| 0.0993 | 0.99422 | 1591.5 | −0.0766 | −14.6 | 27.9 | 0.6010 | 0.88693 | 1392.4 | −0.2384 | −36.7 | 40.4 |
| 0.1484 | 0.98303 | 1568.6 | −0.1080 | −20.2 | 36.2 | 0.6509 | 0.87702 | 1374.7 | −0.2303 | −33.5 | 35.1 |
| 0.2005 | 0.97151 | 1545.3 | −0.1553 | −25.5 | 42.5 | 0.7004 | 0.86740 | 1359.7 | −0.2278 | −31.4 | 31.5 |
| 0.2513 | 0.96012 | 1522.7 | −0.1662 | −29.1 | 45.8 | 0.7512 | 0.85758 | 1344.6 | −0.2118 | −28.2 | 27.2 |
| 0.3013 | 0.94925 | 1501.6 | −0.1913 | −32.3 | 47.9 | 0.8004 | 0.84810 | 1330.3 | −0.1805 | −24.4 | 22.6 |
| 0.3485 | 0.93915 | 1482.4 | −0.2141 | −34.5 | 48.5 | 0.8506 | 0.83865 | 1316.3 | −0.1563 | −19.9 | 17.7 |
| 0.4009 | 0.92800 | 1461.8 | −0.2259 | −36.1 | 48.0 | 0.8999 | 0.82941 | 1302.7 | −0.1181 | −14.5 | 12.4 |
| 0.4488 | 0.91793 | 1443.8 | −0.2310 | −36.9 | 46.8 | 0.9501 | 0.82000 | 1289.0 | −0.0585 | −7.8 | 6.5 |
| 0.5009 | 0.90724 | 1425.0 | −0.2450 | −37.2 | 44.7 | | | | | | |

[a] Standard uncertainties ($u$): $u(T)$ = 0.01 K; $u(p)$ = 1 kPa ; $u(x_1)$ = 0.0005; $u(c)$ = 0.2 m·s$^{-1}$; $u(V_m^E)$ = 0.010·$|V_m^E|_{max}$ + 0.005 cm$^3$·mol$^{-1}$; $u(c^E)$ = 0.4 m·s$^{-1}$. Relative standard uncertainties ($u_r$): $u_r(\rho)$ = 0.0008.

Table 4

Volume fractions of alkan-1-ol ($\phi_1$), relative permittivities at frequency $f = 1$ MHz ($\varepsilon_r$) and excess relative permittivities at $f = 1$ MHz ($\varepsilon_r^E$) of alkan-1-ol (1) + aniline (2) liquid mixtures as functions of the mole fraction of the alkan-1-ol ($x_1$), at temperature $T$ and pressure $p = 0.1$ MPa.

a

| $x_1$ | $\phi_1$ | $\varepsilon_r$ | $\varepsilon_r^E$ | $x_1$ | $\phi_1$ | $\varepsilon_r$ | $\varepsilon_r^E$ |
|---|---|---|---|---|---|---|---|
| methanol (1) + aniline (2) ; $T$/K = 293.15 | | | | | | | |
| 0.0502 | 0.0229 | 7.575 | –0.167 | 0.5508 | 0.3525 | 15.591 | –0.866 |
| 0.1006 | 0.0473 | 8.049 | –0.339 | 0.5977 | 0.3974 | 16.804 | –0.840 |
| 0.1516 | 0.0735 | 8.590 | –0.490 | 0.6507 | 0.4526 | 18.355 | –0.748 |
| 0.2003 | 0.1001 | 9.158 | –0.626 | 0.6971 | 0.5053 | 19.798 | –0.699 |
| 0.2510 | 0.1295 | 9.846 | –0.715 | 0.7501 | 0.5713 | 21.646 | –0.596 |
| 0.3024 | 0.1614 | 10.582 | –0.822 | 0.7997 | 0.6393 | 23.532 | –0.507 |
| 0.3489 | 0.1922 | 11.371 | –0.848 | 0.8497 | 0.7151 | 25.655 | –0.389 |
| 0.3976 | 0.2266 | 12.234 | –0.894 | 0.8992 | 0.7984 | 27.960 | –0.286 |
| 0.4504 | 0.2667 | 13.318 | –0.870 | 0.9500 | 0.8940 | 30.590 | –0.183 |
| 0.4980 | 0.3057 | 14.331 | –0.888 | | | | |
| methanol (1) + aniline (2) ; $T$/K = 298.15 | | | | | | | |
| 0.0502 | 0.0230 | 7.421 | –0.172 | 0.5508 | 0.3530 | 15.126 | –0.922 |
| 0.1006 | 0.0474 | 7.872 | –0.346 | 0.5977 | 0.3979 | 16.295 | –0.903 |
| 0.1516 | 0.0736 | 8.391 | –0.499 | 0.6507 | 0.4532 | 17.801 | –0.814 |
| 0.2003 | 0.1003 | 8.934 | –0.640 | 0.6971 | 0.5059 | 19.188 | –0.777 |
| 0.2510 | 0.1297 | 9.592 | –0.735 | 0.7501 | 0.5718 | 20.997 | –0.657 |
| 0.3024 | 0.1617 | 10.297 | –0.850 | 0.7997 | 0.6398 | 22.825 | –0.571 |
| 0.3489 | 0.1925 | 11.059 | –0.877 | 0.8497 | 0.7155 | 24.895 | –0.440 |
| 0.3976 | 0.2270 | 11.888 | –0.932 | 0.8992 | 0.7987 | 27.127 | –0.340 |
| 0.4504 | 0.2672 | 12.929 | –0.921 | 0.9500 | 0.8942 | 29.693 | –0.220 |
| 0.4980 | 0.3062 | 13.908 | –0.941 | | | | |
| methanol (1) + aniline (2) ; $T$/K = 303.15 | | | | | | | |
| 0.0502 | 0.0230 | 7.273 | –0.174 | 0.5508 | 0.3532 | 14.668 | –0.970 |
| 0.1006 | 0.0475 | 7.702 | –0.352 | 0.5977 | 0.3982 | 15.814 | –0.941 |
| 0.1516 | 0.0737 | 8.196 | –0.508 | 0.6507 | 0.4535 | 17.248 | –0.878 |
| 0.2003 | 0.1004 | 8.718 | –0.649 | 0.6971 | 0.5062 | 18.600 | –0.834 |
| 0.2510 | 0.1299 | 9.344 | –0.755 | 0.7501 | 0.5721 | 20.353 | –0.716 |
| 0.3024 | 0.1618 | 10.022 | –0.868 | 0.7997 | 0.6400 | 22.136 | –0.617 |
| 0.3489 | 0.1927 | 10.751 | –0.906 | 0.8497 | 0.7157 | 24.131 | –0.500 |
| 0.3976 | 0.2272 | 11.554 | –0.958 | 0.8992 | 0.7989 | 26.324 | –0.371 |
| 0.4504 | 0.2674 | 12.548 | –0.962 | 0.9500 | 0.8943 | 28.809 | –0.253 |
| 0.4980 | 0.3064 | 13.497 | –0.980 | | | | |
| propan-1-ol (1) + aniline (2) ; $T$/K = 293.15 | | | | | | | |
| 0.0498 | 0.0412 | 7.463 | –0.254 | 0.5494 | 0.5001 | 12.311 | –1.871 |
| 0.0994 | 0.0830 | 7.802 | –0.504 | 0.5998 | 0.5515 | 13.043 | –1.864 |
| 0.1508 | 0.1272 | 8.175 | –0.754 | 0.6507 | 0.6045 | 13.868 | –1.785 |
| 0.2002 | 0.1704 | 8.558 | –0.980 | 0.7005 | 0.6574 | 14.724 | –1.674 |
| 0.2478 | 0.2128 | 8.952 | –1.183 | 0.7505 | 0.7117 | 15.669 | –1.494 |
| 0.3001 | 0.2602 | 9.421 | −1.382 | 0.8004 | 0.7669 | 16.650 | −1.291 |
| 0.3518 | 0.3081 | 9.926 | −1.552 | 0.8498 | 0.8228 | 17.713 | −1.016 |
| 0.4015 | 0.3550 | 10.446 | −1.692 | 0.8998 | 0.8805 | 18.844 | −0.697 |
| 0.4500 | 0.4017 | 11.014 | −1.782 | 0.9502 | 0.9400 | 20.026 | −0.354 |
| 0.5003 | 0.4510 | 11.632 | −1.859 | | | | |
| | | propan-1-ol (1) + aniline (2) ; $T/K = 298.15$ | | | | | |
| 0.0498 | 0.0413 | 7.315 | −0.248 | 0.5494 | 0.5003 | 11.933 | −1.846 |
| 0.0994 | 0.0831 | 7.636 | −0.493 | 0.5998 | 0.5517 | 12.636 | −1.839 |
| 0.1508 | 0.1273 | 7.992 | −0.736 | 0.6507 | 0.6047 | 13.424 | −1.768 |
| 0.2002 | 0.1705 | 8.357 | −0.956 | 0.7005 | 0.6576 | 14.246 | −1.663 |
| 0.2478 | 0.2129 | 8.732 | −1.155 | 0.7505 | 0.7118 | 15.153 | −1.489 |
| 0.3001 | 0.2604 | 9.178 | −1.352 | 0.8004 | 0.7671 | 16.100 | −1.291 |
| 0.3518 | 0.3083 | 9.659 | −1.520 | 0.8498 | 0.8229 | 17.125 | −1.022 |
| 0.4015 | 0.3552 | 10.157 | −1.657 | 0.8998 | 0.8806 | 18.222 | −0.706 |
| 0.4500 | 0.4019 | 10.693 | −1.753 | 0.9502 | 0.9400 | 19.375 | −0.358 |
| 0.5003 | 0.4512 | 11.289 | −1.825 | | | | |
| | | propan-1-ol (1) + aniline (2) ; $T/K = 303.15$ | | | | | |
| 0.0498 | 0.0413 | 7.171 | −0.242 | 0.5494 | 0.5005 | 11.564 | −1.817 |
| 0.0994 | 0.0832 | 7.476 | −0.481 | 0.5998 | 0.5519 | 12.243 | −1.806 |
| 0.1508 | 0.1273 | 7.815 | −0.716 | 0.6507 | 0.6049 | 12.987 | −1.751 |
| 0.2002 | 0.1706 | 8.164 | −0.929 | 0.7005 | 0.6578 | 13.783 | −1.642 |
| 0.2478 | 0.2130 | 8.518 | −1.126 | 0.7505 | 0.7120 | 14.645 | −1.485 |
| 0.3001 | 0.2606 | 8.944 | −1.319 | 0.8004 | 0.7672 | 15.565 | −1.282 |
| 0.3518 | 0.3084 | 9.399 | −1.485 | 0.8498 | 0.8230 | 16.550 | −1.023 |
| 0.4015 | 0.3554 | 9.878 | −1.617 | 0.8998 | 0.8807 | 17.609 | −0.713 |
| 0.4500 | 0.4020 | 10.382 | −1.719 | 0.9502 | 0.9400 | 18.728 | −0.365 |
| 0.5003 | 0.4514 | 10.956 | −1.787 | | | | |
| | | pentan-1-ol (1) + aniline (2) ; $T/K = 293.15$ | | | | | |
| 0.0574 | 0.0674 | 7.377 | −0.338 | 0.5470 | 0.5891 | 9.890 | −2.318 |
| 0.0993 | 0.1157 | 7.548 | −0.583 | 0.6010 | 0.6414 | 10.287 | −2.371 |
| 0.1484 | 0.1714 | 7.749 | −0.862 | 0.6509 | 0.6888 | 10.790 | −2.276 |
| 0.2005 | 0.2294 | 7.972 | −1.138 | 0.7004 | 0.7351 | 11.278 | −2.187 |
| 0.2513 | 0.2849 | 8.188 | −1.400 | 0.7512 | 0.7819 | 11.873 | −1.995 |
| 0.3013 | 0.3386 | 8.415 | −1.636 | 0.8004 | 0.8264 | 12.501 | −1.750 |
| 0.3485 | 0.3884 | 8.651 | −1.829 | 0.8506 | 0.8711 | 13.219 | −1.417 |
| 0.4009 | 0.4427 | 8.929 | −2.018 | 0.8999 | 0.9143 | 13.979 | −1.029 |
| 0.4488 | 0.4915 | 9.215 | −2.152 | 0.9501 | 0.9576 | 14.848 | −0.533 |
| 0.5009 | 0.5437 | 9.550 | −2.267 | | | | |
| | | pentan-1-ol (1) + aniline (2) ; $T/K = 298.15$ | | | | | |
| 0.0574 | 0.0674 | 7.230 | −0.324 | 0.5470 | 0.5892 | 9.595 | −2.216 |
| 0.0993 | 0.1158 | 7.392 | −0.557 | 0.6010 | 0.6414 | 9.966 | −2.271 |
| 0.1484 | 0.1715 | 7.581 | −0.822 | 0.6509 | 0.6889 | 10.440 | −2.184 |
| 0.2005 | 0.2295 | 7.793 | −1.083 | 0.7004 | 0.7352 | 10.901 | −2.101 |
| 0.2513 | 0.2850 | 7.996 | −1.333 | 0.7512 | 0.7819 | 11.460 | −1.923 |
| 0.3013 | 0.3387 | 8.210 | −1.557 | 0.8004 | 0.8265 | 12.059 | −1.688 |
| 0.3485 | 0.3885 | 8.432 | −1.741 | 0.8506 | 0.8712 | 12.736 | −1.375 |
| 0.4009 | 0.4428 | 8.694 | −1.922 | 0.8999 | 0.9144 | 13.465 | −0.999 |
| 0.4488 | 0.4916 | 8.960 | −2.054 | 0.9501 | 0.9576 | 14.295 | −0.521 |
| 0.5009 | 0.5438 | 9.275 | −2.165 | | | | |
| pentan-1-ol (1) + aniline (2) ; $T$/K = 303.15 | | | | | | | |
| 0.0574 | 0.0675 | 7.088 | −0.308 | 0.5470 | 0.5892 | 9.305 | −2.114 |
| 0.0993 | 0.1158 | 7.242 | −0.527 | 0.6010 | 0.6415 | 9.657 | −2.165 |
| 0.1484 | 0.1715 | 7.418 | −0.780 | 0.6509 | 0.6889 | 10.100 | −2.087 |
| 0.2005 | 0.2295 | 7.619 | −1.026 | 0.7004 | 0.7352 | 10.538 | −2.006 |
| 0.2513 | 0.2850 | 7.808 | −1.265 | 0.7512 | 0.7820 | 11.058 | −1.847 |
| 0.3013 | 0.3387 | 8.013 | −1.474 | 0.8004 | 0.8265 | 11.628 | −1.620 |
| 0.3485 | 0.3885 | 8.218 | −1.653 | 0.8506 | 0.8712 | 12.263 | −1.330 |
| 0.4009 | 0.4429 | 8.465 | −1.826 | 0.8999 | 0.9144 | 12.962 | −0.964 |
| 0.4488 | 0.4917 | 8.714 | −1.953 | 0.9501 | 0.9577 | 13.752 | −0.508 |
| 0.5009 | 0.5438 | 9.012 | −2.057 | | | | |

[a] Standard uncertainties ($u$): $u(T)$ = 0.02 K; $u(p)$ = 1 kPa; $u(f)$ = 20 Hz; $u(x_1)$ = 0.0005; $u(\phi_1)$ = 0.004. Relative standard uncertainty ($u_r$): $u_r(\varepsilon_r)$ = 0.003. Relative expanded uncertainty at 0.95 confidence level ($U_r$): $U_r(\varepsilon_r^E)$ = 0.03.

Table 5

Volume fractions of alkan-1-ol ($\phi_1$), refractive indices at the sodium D-line ($n_D$) and excess refractive indices at the sodium D-line ($n_D^E$) of alkan-1-ol (1) + aniline (2) liquid mixtures as functions of the mole fraction of the alkan-1-ol, $x_1$, at temperature $T$ and pressure $p = 0.1$ MPa. [a]

| $x_1$ | $\phi_1$ | $n_D$ | $10^5 n_D^E$ | $x_1$ | $\phi_1$ | $n_D$ | $10^5 n_D^E$ |
|---|---|---|---|---|---|---|---|
| methanol (1) + aniline (2) ; $T$/K = 293.15 | | | | | | | |
| 0.0519 | 0.0237 | 1.58144 | 51 | 0.5457 | 0.3478 | 1.50271 | 86 |
| 0.0974 | 0.0457 | 1.57619 | 50 | 0.6459 | 0.4474 | 1.47749 | 79 |
| 0.1546 | 0.0751 | 1.56942 | 76 | 0.7501 | 0.5713 | 1.44514 | 34 |
| 0.2524 | 0.1303 | 1.55609 | 73 | 0.7981 | 0.6370 | 1.42737 | −22 |
| 0.3471 | 0.1909 | 1.54138 | 74 | 0.8488 | 0.7136 | 1.40674 | −53 |
| 0.4539 | 0.2695 | 1.52214 | 81 | 0.8989 | 0.7978 | 1.38391 | −67 |
| 0.5022 | 0.3093 | 1.51233 | 87 | 0.9472 | 0.8884 | 1.35897 | −78 |
| methanol (1) + aniline (2) ; $T$/K = 298.15 | | | | | | | |
| 0.0519 | 0.0238 | 1.57872 | 58 | 0.5457 | 0.3483 | 1.50113 | 198 |
| 0.0974 | 0.0458 | 1.57367 | 76 | 0.6459 | 0.4480 | 1.47586 | 182 |
| 0.1546 | 0.0752 | 1.56691 | 101 | 0.7501 | 0.5718 | 1.44339 | 115 |
| 0.2021 | 0.1013 | 1.56081 | 117 | 0.7981 | 0.6375 | 1.42550 | 42 |
| 0.2524 | 0.1306 | 1.55384 | 125 | 0.8488 | 0.7141 | 1.40485 | 5 |
| 0.3471 | 0.1913 | 1.53929 | 141 | 0.8989 | 0.7982 | 1.38208 | −12 |
| 0.3988 | 0.2279 | 1.53051 | 157 | 0.9472 | 0.8886 | 1.35694 | −54 |
| 0.4539 | 0.2699 | 1.52030 | 169 | | | | |
| methanol (1) + aniline (2) ; $T$/K = 303.15 | | | | | | | |
| 0.0519 | 0.0238 | 1.57630 | 84 | 0.6459 | 0.4482 | 1.47401 | 246 |
| 0.0974 | 0.0459 | 1.57114 | 92 | 0.7501 | 0.5721 | 1.44193 | 213 |
| 0.1546 | 0.0753 | 1.56449 | 127 | 0.7981 | 0.6378 | 1.42425 | 157 |
| 0.2021 | 0.1014 | 1.55839 | 141 | 0.8488 | 0.7143 | 1.40373 | 125 |
| 0.3009 | 0.1609 | 1.54421 | 155 | 0.8989 | 0.7984 | 1.38083 | 90 |
| 0.3988 | 0.2281 | 1.52834 | 201 | 0.9472 | 0.8888 | 1.35538 | 11 |
| 0.5022 | 0.3100 | 1.50847 | 229 | | | | |
| propan-1-ol (1) + aniline (2) ; $T$/K = 293.15 | | | | | | | |
| 0.0548 | 0.0454 | 1.57744 | −53 | 0.6014 | 0.5532 | 1.47692 | −160 |
| 0.1050 | 0.0878 | 1.56918 | −73 | 0.6488 | 0.6025 | 1.46704 | −147 |
| 0.1455 | 0.1226 | 1.56224 | −102 | 0.7530 | 0.7144 | 1.44442 | −110 |
| 0.2543 | 0.2186 | 1.54332 | −145 | 0.8481 | 0.8208 | 1.42262 | −70 |
| 0.3488 | 0.3053 | 1.52617 | −171 | 0.8889 | 0.8678 | 1.41298 | −43 |
| 0.4500 | 0.4017 | 1.50708 | −180 | 0.9521 | 0.9422 | 1.39733 | −23 |
| 0.5527 | 0.5034 | 1.48687 | −170 | | | | |
| propan-1-ol (1) + aniline (2) ; $T$/K = 298.15 | | | | | | | |
| 0.0548 | 0.0454 | 1.57478 | −44 | 0.6014 | 0.5534 | 1.47459 | −150 |
| 0.1050 | 0.0879 | 1.56657 | −60 | 0.6488 | 0.6027 | 1.46478 | −134 |
| 0.1455 | 0.1227 | 1.55968 | −87 | 0.7530 | 0.7146 | 1.44221 | −101 |
| 0.2543 | 0.2188 | 1.54082 | −128 | 0.8481 | 0.8209 | 1.42052 | −60 |
| 0.3488 | 0.3055 | 1.52373 | −154 | 0.8889 | 0.8679 | 1.41080 | −44 |
| 0.4500 | 0.4019 | 1.50466 | −168 | 0.9521 | 0.9423 | 1.39528 | −17 |
| 0.5527 | 0.5036 | 1.48450 | −160 | | | | |
| | | propan-1-ol (1) + aniline (2) ; $T/K$ = 303.15 | | | | | |
| 0.0548 | 0.0455 | 1.57219 | −35 | 0.6014 | 0.5535 | 1.47251 | −120 |
| 0.1050 | 0.0879 | 1.56406 | −47 | 0.6488 | 0.6029 | 1.46267 | −107 |
| 0.1455 | 0.1228 | 1.55720 | −70 | 0.7530 | 0.7147 | 1.44013 | −80 |
| 0.2543 | 0.2189 | 1.53848 | −104 | 0.8481 | 0.8211 | 1.41845 | −43 |
| 0.3488 | 0.3056 | 1.52146 | −128 | 0.8889 | 0.8680 | 1.40876 | −29 |
| 0.4500 | 0.4020 | 1.50249 | −137 | 0.9521 | 0.9423 | 1.39320 | −14 |
| 0.5527 | 0.5038 | 1.48239 | −128 | | | | |
| | | pentan-1-ol (1) + aniline (2) ; $T/K$ = 293.15 | | | | | |
| 0.0528 | 0.0621 | 1.57551 | −66 | 0.5534 | 0.5953 | 1.48081 | −318 |
| 0.1002 | 0.1168 | 1.56558 | −138 | 0.5969 | 0.6374 | 1.47341 | −305 |
| 0.1465 | 0.1693 | 1.55630 | −178 | 0.6495 | 0.6875 | 1.46464 | −282 |
| 0.1979 | 0.2266 | 1.54608 | −224 | 0.6989 | 0.7337 | 1.45645 | −266 |
| 0.2508 | 0.2844 | 1.53575 | −266 | 0.7523 | 0.7829 | 1.44793 | −223 |
| 0.2961 | 0.3331 | 1.52722 | −279 | 0.7980 | 0.8243 | 1.44063 | −196 |
| 0.3543 | 0.3945 | 1.51621 | −315 | 0.8451 | 0.8663 | 1.43337 | −149 |
| 0.4008 | 0.4426 | 1.50770 | −326 | 0.8996 | 0.9141 | 1.42504 | −99 |
| 0.4542 | 0.4970 | 1.49814 | −327 | 0.9487 | 0.9564 | 1.41763 | −53 |
| 0.4966 | 0.5394 | 1.49066 | −326 | | | | |
| | | pentan-1-ol (1) + aniline (2) ; $T/K$ = 298.15 | | | | | |
| 0.0528 | 0.0621 | 1.57276 | −68 | 0.5534 | 0.5954 | 1.47848 | −315 |
| 0.1002 | 0.1168 | 1.56294 | −133 | 0.5969 | 0.6375 | 1.47113 | −301 |
| 0.1465 | 0.1693 | 1.55362 | −180 | 0.6495 | 0.6876 | 1.46240 | −278 |
| 0.1979 | 0.2266 | 1.54353 | −217 | 0.6989 | 0.7338 | 1.45431 | −256 |
| 0.2508 | 0.2845 | 1.53314 | −268 | 0.7523 | 0.7829 | 1.44579 | −219 |
| 0.2961 | 0.3331 | 1.52455 | −293 | 0.7980 | 0.8243 | 1.43851 | −193 |
| 0.3543 | 0.3945 | 1.51372 | −315 | 0.8451 | 0.8663 | 1.43129 | −146 |
| 0.4008 | 0.4427 | 1.50515 | −334 | 0.8996 | 0.9141 | 1.42301 | −94 |
| 0.4542 | 0.4971 | 1.49568 | −330 | 0.9487 | 0.9565 | 1.41560 | −50 |
| 0.4966 | 0.5395 | 1.48826 | −326 | | | | |
| | | pentan-1-ol (1) + aniline (2) ; $T/K$ = 303.15 | | | | | |
| 0.0528 | 0.0621 | 1.57005 | −73 | 0.5534 | 0.5954 | 1.47615 | −317 |
| 0.1002 | 0.1168 | 1.56032 | −133 | 0.5969 | 0.6375 | 1.46881 | −304 |
| 0.1465 | 0.1694 | 1.55101 | −180 | 0.6495 | 0.6876 | 1.46008 | −284 |
| 0.1979 | 0.2266 | 1.54089 | −226 | 0.6989 | 0.7338 | 1.45198 | −266 |
| 0.2508 | 0.2845 | 1.53061 | −269 | 0.7523 | 0.7830 | 1.44353 | −224 |
| 0.2961 | 0.3332 | 1.52205 | −292 | 0.7980 | 0.8243 | 1.43634 | −194 |
| 0.3543 | 0.3946 | 1.51128 | −312 | 0.8451 | 0.8663 | 1.42912 | −150 |
| 0.4008 | 0.4427 | 1.50275 | −332 | 0.8996 | 0.9141 | 1.42089 | −96 |
| 0.4542 | 0.4971 | 1.49329 | −331 | 0.9487 | 0.9565 | 1.41352 | −51 |
| 0.4966 | 0.5395 | 1.48594 | −323 | | | | |

[a] Standard uncertainties ($u$): $u(T)$ = 0.02 K; $u(p)$ = 1 kPa; $u(x_1)$ = 0.0005; $u(\phi_1)$ = 0.004, $u(n_D)$ = 0.00008. Expanded uncertainty at 0.95 confidence level ($U$): $U(n_D^E)$ = 0.0002.

Table 6

Volume fraction of alkan-1-ol ($\phi_1$) and temperature derivative of the excess relative permittivity at frequency $f = 1$ MHz ($(\partial \varepsilon_r^E/\partial T)_p$) of alkan-1-ol (1) + aniline (2) liquid mixtures as functions of the mole fraction of the alkan-1-ol ($x_1$) at temperature $T$ and pressure $p = 0.1$ MPa.[a]

| $x_1$ | $\phi_1$ | $(\partial \varepsilon_r^E/\partial T)_p$ | $x_1$ | $\phi_1$ | $(\partial \varepsilon_r^E/\partial T)_p$ |
|---|---|---|---|---|---|
| methanol (1) + aniline (2) ; $T/K = 298.15$ ||||||
| 0.0502 | 0.0230 | –0.0007 | 0.5508 | 0.3530 | –0.0104 |
| 0.1006 | 0.0474 | –0.0013 | 0.5977 | 0.3979 | –0.0101 |
| 0.1516 | 0.0736 | –0.0018 | 0.6507 | 0.4532 | –0.0130 |
| 0.2003 | 0.1003 | –0.0023 | 0.6971 | 0.5059 | –0.0135 |
| 0.2510 | 0.1297 | –0.0040 | 0.7501 | 0.5718 | –0.0120 |
| 0.3024 | 0.1617 | –0.0046 | 0.7997 | 0.6398 | –0.0110 |
| 0.3489 | 0.1925 | –0.0058 | 0.8497 | 0.7155 | –0.0111 |
| 0.3976 | 0.2270 | –0.0064 | 0.8992 | 0.7987 | –0.0085 |
| 0.4504 | 0.2672 | –0.0092 | 0.9500 | 0.8942 | –0.0070 |
| 0.4980 | 0.3062 | –0.0092 | | | |
| propan-1-ol (1) + aniline (2) ; $T/K = 298.15$ ||||||
| 0.0498 | 0.0413 | 0.0012 | 0.5494 | 0.5003 | 0.0054 |
| 0.0994 | 0.0831 | 0.0023 | 0.5998 | 0.5517 | 0.0058 |
| 0.1508 | 0.1273 | 0.0038 | 0.6507 | 0.6047 | 0.0034 |
| 0.2002 | 0.1705 | 0.0051 | 0.7005 | 0.6576 | 0.0032 |
| 0.2478 | 0.2129 | 0.0057 | 0.7505 | 0.7118 | 0.0009 |
| 0.3001 | 0.2604 | 0.0063 | 0.8004 | 0.7671 | 0.0009 |
| 0.3518 | 0.3083 | 0.0067 | 0.8498 | 0.8229 | –0.0007 |
| 0.4015 | 0.3552 | 0.0075 | 0.8998 | 0.8806 | –0.0016 |
| 0.4500 | 0.4019 | 0.0063 | 0.9502 | 0.9400 | –0.0011 |
| 0.5003 | 0.4512 | 0.0072 | | | |
| pentan-1-ol (1) + aniline (2) ; $T/K = 298.15$ ||||||
| 0.0574 | 0.0674 | 0.0030 | 0.5470 | 0.5892 | 0.0204 |
| 0.0993 | 0.1158 | 0.0056 | 0.6010 | 0.6414 | 0.0206 |
| 0.1484 | 0.1715 | 0.0082 | 0.6509 | 0.6889 | 0.0189 |
| 0.2005 | 0.2295 | 0.0112 | 0.7004 | 0.7352 | 0.0181 |
| 0.2513 | 0.2850 | 0.0135 | 0.7512 | 0.7819 | 0.0148 |
| 0.3013 | 0.3387 | 0.0162 | 0.8004 | 0.8265 | 0.0130 |
| 0.3485 | 0.3885 | 0.0176 | 0.8506 | 0.8712 | 0.0087 |
| 0.4009 | 0.4428 | 0.0192 | 0.8999 | 0.9144 | 0.0065 |
| 0.4488 | 0.4916 | 0.0199 | 0.9501 | 0.9576 | 0.0025 |
| 0.5009 | 0.5438 | 0.0210 | | | |

[a] Standard uncertainties ($u$): $u(T) = 0.02$ K; $u(p) = 1$ kPa; $u(f) = 20$ Hz; $u(x_1) = 0.0005$; $u(\phi_1) = 0.004$; $u\left[(\partial \varepsilon_r^E/\partial T)_p\right] = 0.0008$ K$^{-1}$.

Table 7

Coefficients $A_i$ and standard deviations, $\sigma(F^E)$ (equation (12)), for the representation of $F^E$ at temperature $T$ and pressure $p = 0.1$ MPa for alkan-1-ol (1) + aniline liquid mixtures by equation (11).

| Property $F^E$ | alkan-1-ol | $T$/K | $A_0$ | $A_1$ | $A_2$ | $A_3$ | $A_4$ | $A_5$ | $A_6$ | $\sigma(F^E)$ |
|---|---|---|---|---|---|---|---|---|---|---|
| $V_m^E$ /cm$^3\cdot$mol$^{-1}$ | methanol | 298.15 | −3.607 | −1.39 | −0.75 | −0.59 | −0.4 | | | 0.003 |
| | propan-1-ol | 298.15 | −2.370 | −0.73 | −0.49 | −0.33 | | | | 0.004 |
| | pentan-1-ol | 298.15 | −0.963 | −0.21 | −0.19 | | | | | 0.005 |
| $\kappa_S^E$ /TPa$^{-1}$ | methanol | 298.15 | −438.7 | −434 | −344 | −201 | −105 | −186 | −160 | 0.14 |
| | propan-1-ol | 298.15 | −289.9 | −125 | −63 | −48 | −33 | | | 0.15 |
| | pentan-1-ol | 298.15 | −149.0 | 4 | −19 | | | | | 0.3 |
| $c^E$ /m$\cdot$s$^{-1}$ | methanol | 298.15 | 556.0 | 187.4 | 82 | 37 | 34 | | | 0.12 |
| | propan-1-ol | 298.15 | 346.0 | −52.0 | 45 | 9 | 15 | | | 0.13 |
| | pentan-1-ol | 298.15 | 179.1 | −91 | 68 | −30 | | | | 0.4 |
| $\varepsilon_r^E$ | methanol | 293.15 | −3.54 | 0.57 | | | | | | 0.018 |
| | | 298.15 | −3.77 | 0.31 | | | | | | 0.02 |
| | | 303.15 | −3.95 | | | | | | | 0.02 |
| | propan-1-ol | 293.15 | −7.43 | −1.86 | 1.05 | 0.8 | | | | 0.007 |
| | | 298.15 | −7.286 | −1.97 | 0.6 | 0.76 | 0.6 | | | 0.005 |
| | | 303.15 | −7.146 | −2.01 | 0.48 | 0.3 | 0.48 | 0.5 | | 0.0017 |
| | pentan-1-ol | 293.15 | −9.09 | −3.43 | 0.19 | 0.7 | | | | 0.011 |
| | | 298.15 | −8.68 | −3.39 | 0.04 | 0.6 | | | | 0.010 |
| | | 303.15 | −8.25 | −3.31 | −0.09 | 0.5 | | | | 0.009 |
| $10^5 n_D^E$ | methanol | 293.15 | 369 | 207 | −606 | −1822 | | | | 11 |
| | | 298.15 | 762 | 408 | −690 | −1903 | | | | 11 |
| | | 303.15 | 923 | 658 | 89 | −1396 | | | | 12 |
| | propan-1-ol | 293.15 | −694 | 196 | | | | | | 5 |
| | | 298.15 | −657 | 151 | 158 | | | | | 4 |
| | | 303.15 | −535 | 147 | 176 | | | | | 3 |
| | pentan-1-ol | 293.15 | −1305 | 190 | | | | | | 5 |
| | | 298.15 | −1302 | 230 | | | | | | 5 |
| | | 303.15 | −1311 | 212 | | | | | | 4 |
| $(\partial \varepsilon_r^E / \partial T)_p$ /K$^{-1}$ | methanol | 298.15 | −0.037 | −0.049 | −0.027 | | | | | 0.0008 |
| | propan-1-ol | 298.15 | 0.0266 | −0.023 | −0.030 | | | | | 0.0005 |
| | pentan-1-ol | 298.15 | 0.0834 | 0.008 | −0.028 | | | | | 0.0005 |

Table 8

Literature data on density ($\rho$), speed of sound ($c$), refractive index at the sodium D-line ($n_D$) and relative permittivity ($\varepsilon_r$) of alkan-1-ol (1) + aniline at pressure 0.1 MPa and temperature $T$. Symbols: $n$, number of carbon atoms of the alkan-1-ol; $F^E$, excess property ($F = V_m$, $n_D$, $\varepsilon_r$); $\Delta c$, deviation from mole-fraction linearity of the speed of sound.

| Ref. | Year | $n$ | Properties | $T$/K | Comparison with our data |
|---|---|---|---|---|---|
| [20] | 1964 | 1,2 | $\varepsilon_r$ | 293.15 | Lower $\varepsilon_r^E$ ($n = 1$). |
| [21] | 1967 | 1 | $\rho, n_D$ | 293.15, 303.15, 313.15 | Good agreement in $V_m^E$. Large data scattering in $n_D^E$. |
| [22] | 1971 | 1 | $\rho, c$ | 308.15 | Good agreement in $V_m^E$. Good agreement in $\Delta c$. |
| [23] | 1971 | 4 | $\rho, n_D$ | 298.15, 303.15, 308.15, 313.15 | $V_m^E$ agrees with our $n$ variation and is close to ref. [33]. Symmetry of the curve also agrees with ours. $n_D^E$ agrees rather well with our $n$ variation, but large data scattering. |
| [24] | 1984 | 3,4 | $\varepsilon_r$ | 307.15 | Comparable $\varepsilon_r^E$ values ($n = 3$), but large data scattering. |
| [25] | 1993 | 3,4 | $\varepsilon_r$ | 307.15 | $\varepsilon_r^E$ not computable ($\varepsilon_r$ of pure compounds not provided). |
| [26] | 1994 | 1 | $\varepsilon_r$ | 283.15, 293.15, 303.15, 313.15 | Much lower $\varepsilon_r^E$, and large data scattering. |
| [27] | 1994 | 5,6,7,8 | $\varepsilon_r$ | 303.15 | $\varepsilon_r^E$ not computable ($\varepsilon_r$ of pure compounds not provided). |
| [28] | 1999 | 2,3,4,6,7 | $\varepsilon_r$ | 283.15, 293.15, 303.15, 313.15 | Much higher $\varepsilon_r^E$ ($n = 3$), and large data scattering. |
| [29] | 2000 | 3,4,5,6,7,8 | $\varepsilon_r$ | 303.15 | $\varepsilon_r^E$ not computable ($\varepsilon_r$ of pure compounds not provided). |
| [30] | 2000 | 5 | $\rho$ | 293.15 | Good agreement in $V_m^E$. |
| [31] | 2001 | 3 | $\rho$ | 294.15, 298.15, 303.15, 308.15, 313.15 | Good agreement in $V_m^E$. |
| [32] | 2003 | 3 | $\varepsilon_r$ | 293.15, 303.15, 313.15, 323.15 | Comparable $\varepsilon_r^E$ values, but large data scattering. |
| [33] | 2003 | 4 | $\rho$ | 298.15, 303.15, 308.15, 313.15 | $V_m^E$ agrees with our $n$ variation and is close to ref. [23]. Symmetry of the curve does not agree with ours. |
| [34] | 2007 | 1 | $\rho$ | 298.15 | Positive $V_m^E$, abnormally high. |
| [35] | 2007 | 3 | $\rho$ | 293.15, 298.15, 303.15, 308.15, 313.15, 318.15 | Higher $V_m^E$. |
| [36] | 2007 | 3 | $c$ | 293.15, 298.15, 303.15, 308.15, 313.15, 318.15 | Comparable $\Delta c$ values. |
| [37] | 2010 | 3 | $n_D$ | 293.15, 298.15, 303.15, 308.15, 313.15, 318.15 | Lower $n_D^E$, opposite symmetry of the curve. |
| [38] | 2019 | 3,4,5,6 | $\rho, c$ | 303.15, 308.15, 313.15, 318.15 (for $\rho$) 303.15, 313.15 (for $c$) | Abnormally high $V_m^E$ values, positive for $n$ = 4,5,6. Large data scattering in $\Delta c$; values lower than ours for $n$ = 3 and comparable for $n$ = 5. |

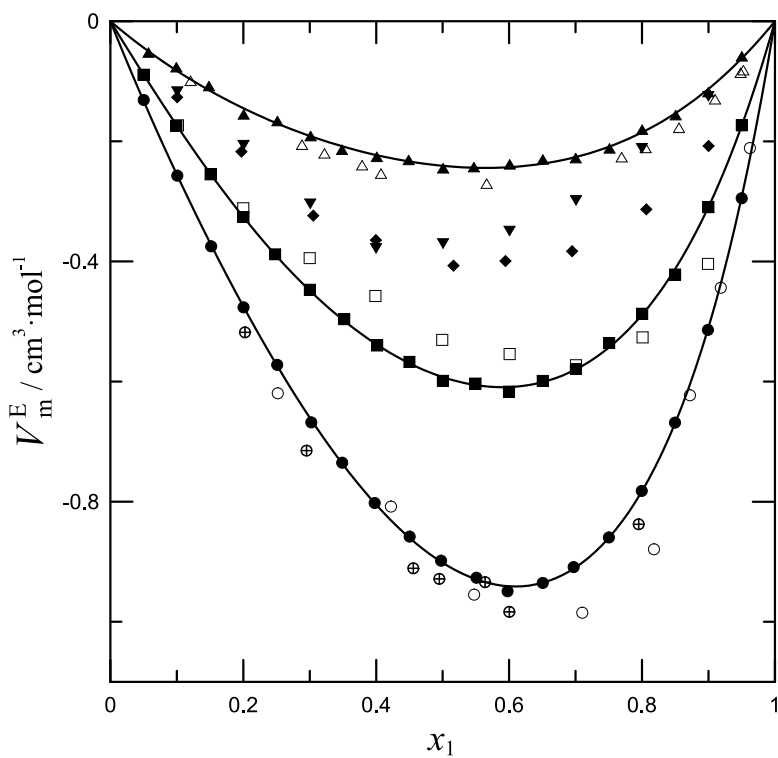

Figure 1. Excess molar volume ($V_m^E$) of alkan-1-ol (1) + aniline (2) liquid mixtures as a function of the alkan-1-ol mole fraction ($x_1$) at 298.15 K and 0.1 MPa. Symbols, experimental values: (●), methanol (this work); (■), propan-1-ol (this work); (▲), pentan-1-ol (this work); (○), methanol ([21], $T$ = 293.15 K); (⊕), methanol ([22], $T$ = 308.15 K); (□), propan-1-ol ([31]); (♦), butan-1-ol ([23]); (▼), butan-1-ol ([33]); (Δ), pentan-1-ol ([30], $T$ = 293.15 K). Solid lines, calculations with equation (11) using the coefficients from Table 7.

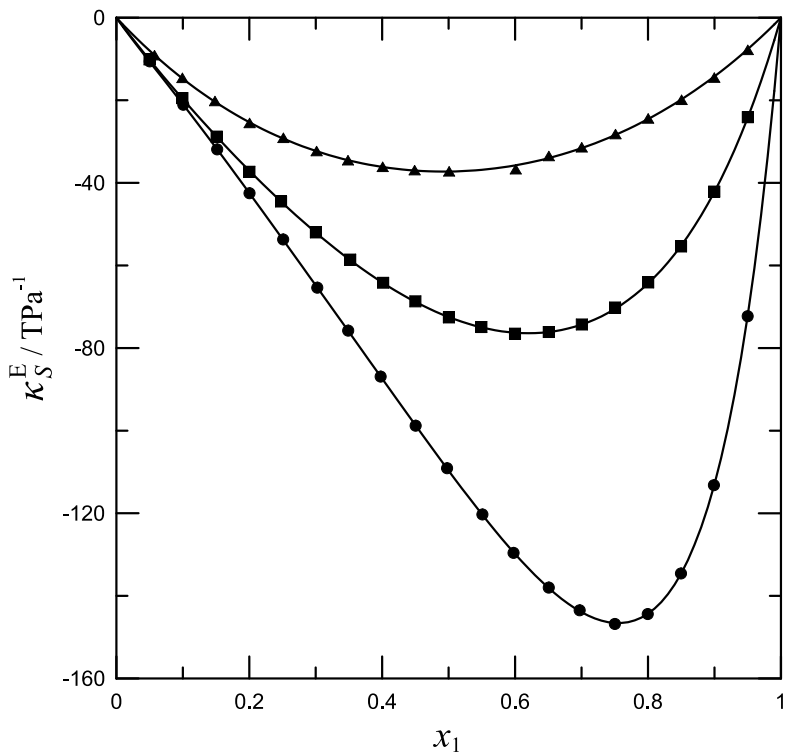

Figure 2. Excess isentropic compressibility ($\kappa_S^E$) of alkan-1-ol (1) + aniline (2) liquid mixtures as a function of the alkan-1-ol mole fraction ($x_1$) at 298.15 K and 0.1 MPa. Full symbols, experimental values (this work): (●), methanol; (■), propan-1-ol; (▲), pentan-1-ol. Solid lines, calculations with equation (11) using the coefficients from Table 7.

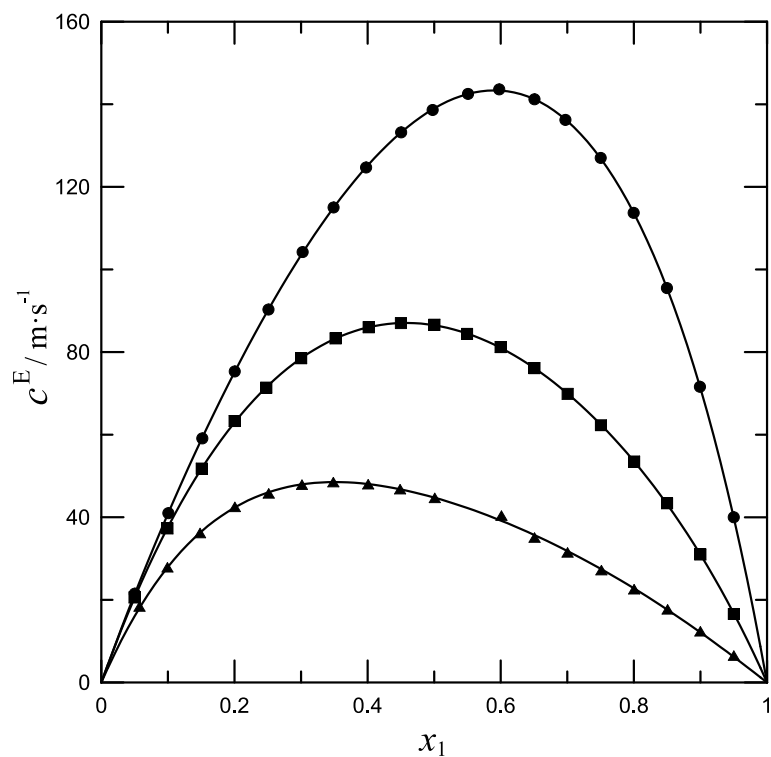

Figure 3. Excess speed of sound ($c^E$) of alkan-1-ol (1) + aniline (2) liquid mixtures as a function of the alkan-1-ol mole fraction ($x_1$) at 298.15 K and 0.1 MPa. Full symbols, experimental values (this work): (●), methanol; (■), propan-1-ol; (▲), pentan-1-ol. Solid lines, calculations with equation (11) using the coefficients from Table 7.

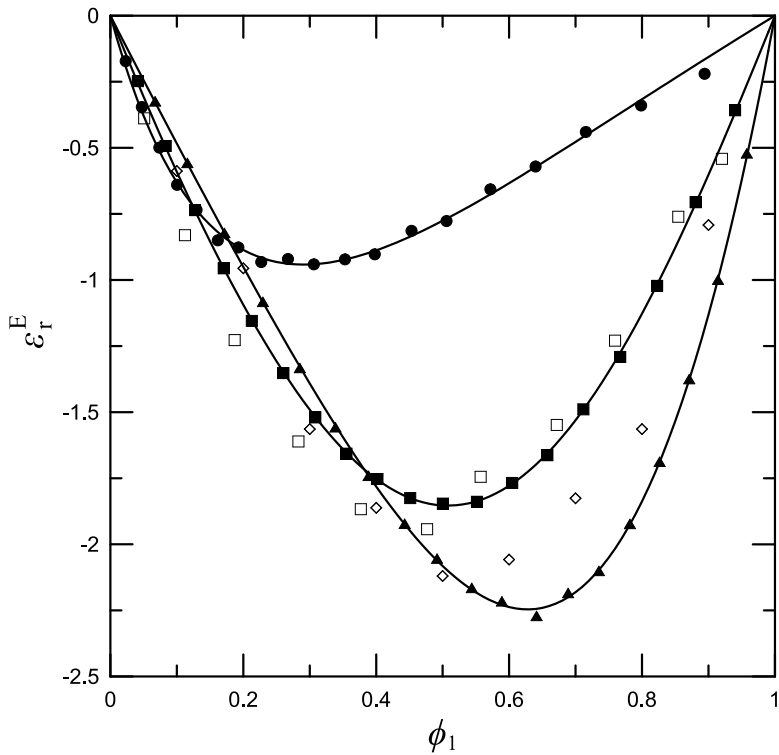

Figure 4. Excess relative permittivity ($\varepsilon_r^E$) of alkan-1-ol (1) + aniline (2) liquid mixtures as a function of the alkan-1-ol volume fraction ($\phi_1$) at 298.15 K, 0.1 MPa, and 1 MHz. Symbols, experimental values: (●), methanol (this work); (■), propan-1-ol (this work); (▲), pentan-1-ol (this work); (□), propan-1-ol ([24], $T$ = 307.15 K); (◊), propan-1-ol ([32], $T$ = 303.15 K). Solid lines, calculations with equation (11) using the coefficients from Table 7.

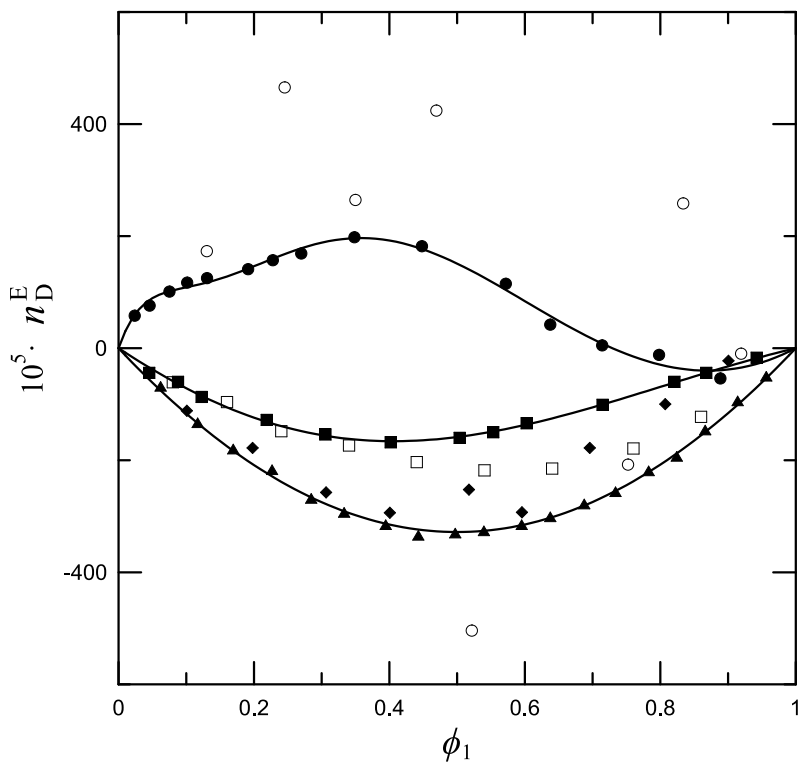

Figure 5. Excess refractive index at the sodium D-line ($n_\text{D}^\text{E}$) of alkan-1-ol (1) + aniline (2) liquid mixtures as a function of the alkan-1-ol volume fraction ($\phi_1$) at 298.15 K and 0.1 MPa. Full symbols, experimental values: (●), methanol (this work); (■), propan-1-ol (this work); (▲), pentan-1-ol (this work); (○), methanol ([21], $T$ = 303.15 K); (□), propan-1-ol ([37]); (♦), butan-1-ol ([23]). Solid lines, calculations with equation (11) using the coefficients from Table 7.

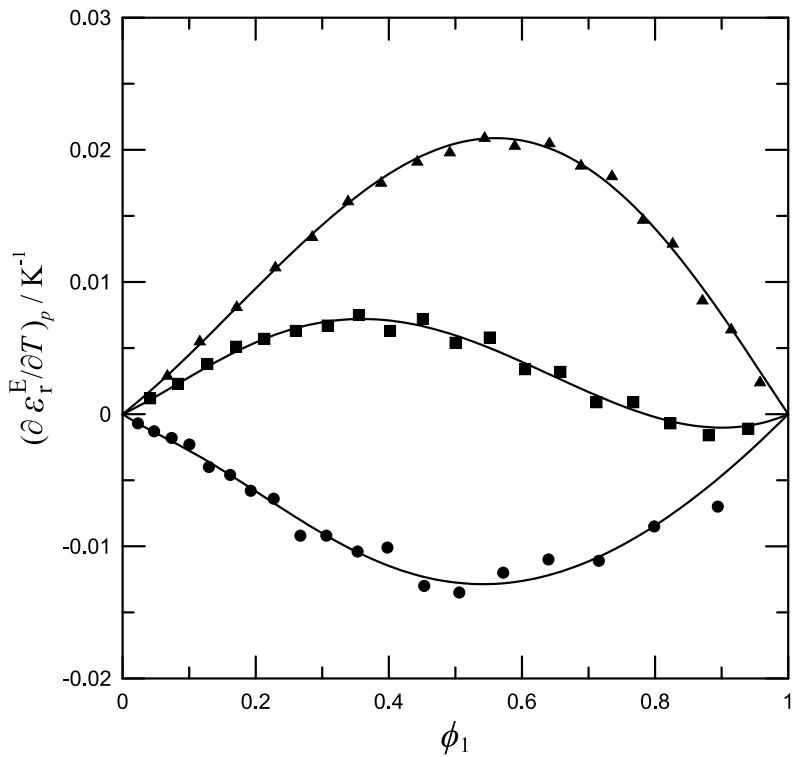

Figure 6. Temperature derivative of the excess relative permittivity ($(\partial \varepsilon_r^E / \partial T)_p$) of alkan-1-ol (1) + aniline (2) liquid mixtures as a function of the alkan-1-ol volume fraction ($\phi_1$) at 298.15 K, 0.1 MPa, and 1 MHz. Full symbols, experimental values (this work): (●), methanol; (■), propan-1-ol; (▲), pentan-1-ol. Solid lines, calculations with equation (11) using the coefficients from Table 7.

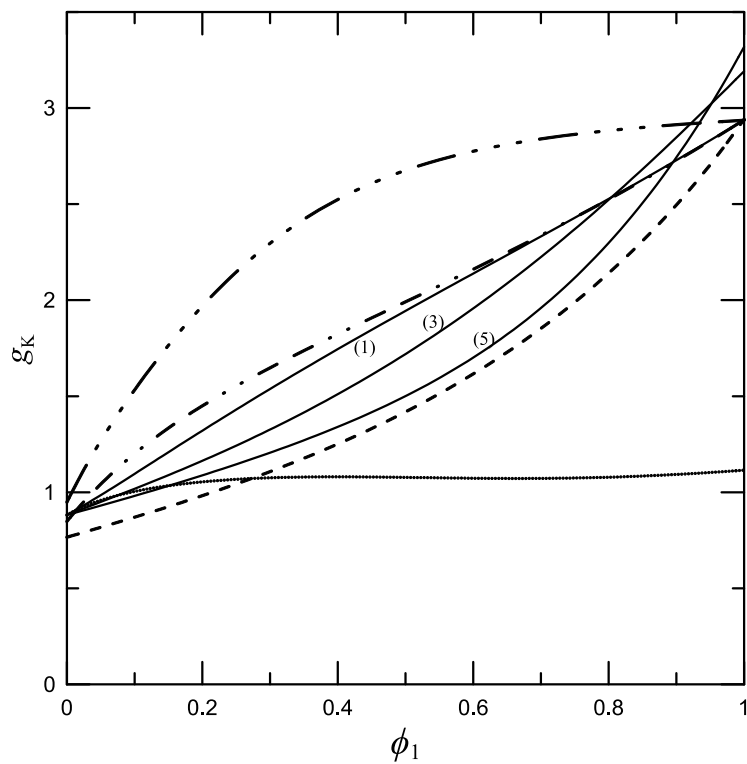

Figure 7. Kirkwood correlation factor ($g_K$) of several (1) + (2) liquid mixtures as a function of the volume fraction of compound (1), $\phi_1$, at 0.1 MPa and 298.15 K: Solid lines, alkan-1-ol (1) + aniline (2) (this work); the numbers in parentheses indicate the number of carbon atoms of the alkan-1-ol. (– – –), methanol (1) + benzonitrile (2) (293.15 K, [20]). (······), *N,N*-dimethylformamide (1) + aniline (2) [78]. (– – ·· – –), methanol (1) + pyridine (2) [68]. (– ··· –), methanol (1) + hexan-1-amine (2) [13].

**Supplementary material**

**Density, speed of sound, refractive index and relative permittivity of methanol, propan-1-ol or pentan-1-ol + aniline liquid mixtures. Application of the Kirkwood-Fröhlich model**

Fernando Hevia*[,a], Víctor Alonso[b], Juan Antonio González[c], Luis Felipe Sanz[c], Isaías García de la Fuente[c], José Carlos Cobos[c]

[a] Université Clermont Auvergne, CNRS. Institut de Chimie de Clermont-Ferrand. F-63000, Clermont-Ferrand, France

[b] Departamento de Física Aplicada. EIFAB. Campus Duques de Soria. Universidad de Valladolid. C/ Universidad s.n. 42004 Soria, Spain.

[c] G.E.T.E.F., Departamento de Física Aplicada. Facultad de Ciencias. Universidad de Valladolid. Paseo de Belén, 7, 47011 Valladolid, Spain.

*e-mail: fernando.hevia@termo.uva.es

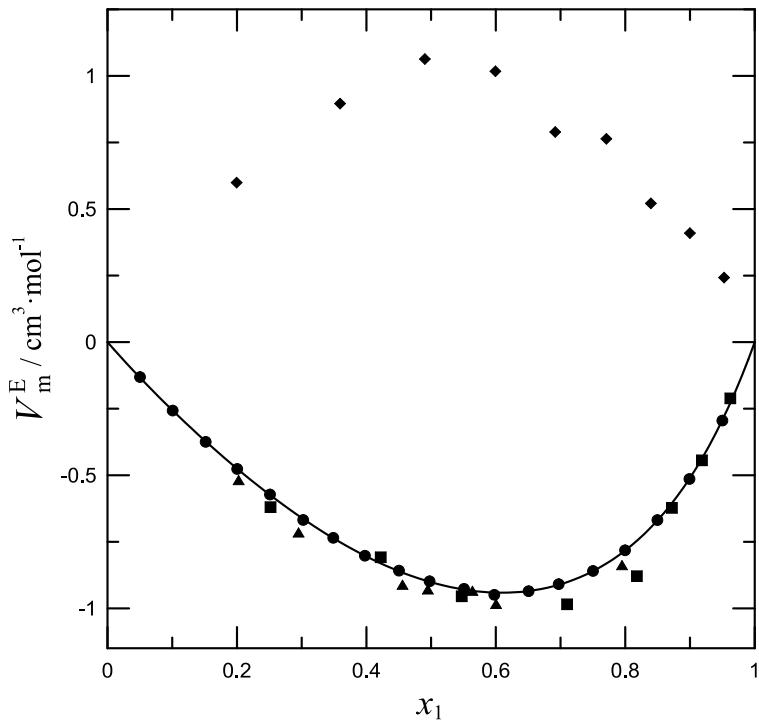

Figure S1. Excess molar volume ($V_m^E$) of the methanol (1) + aniline (2) liquid mixture as a function of the alkan-1-ol mole fraction ($x_1$) at temperature $T$ and 0.1 MPa. Full symbols, experimental values: (●), this work, $T$ = 298.15 K; (■), ref. [21], $T$ = 293.15 K; (▲), ref. [22], $T$ = 308.15 K; (♦) ref. [34], $T$ = 298.15 K. Solid line, calculations with equation (11) using the coefficients from Table 7.

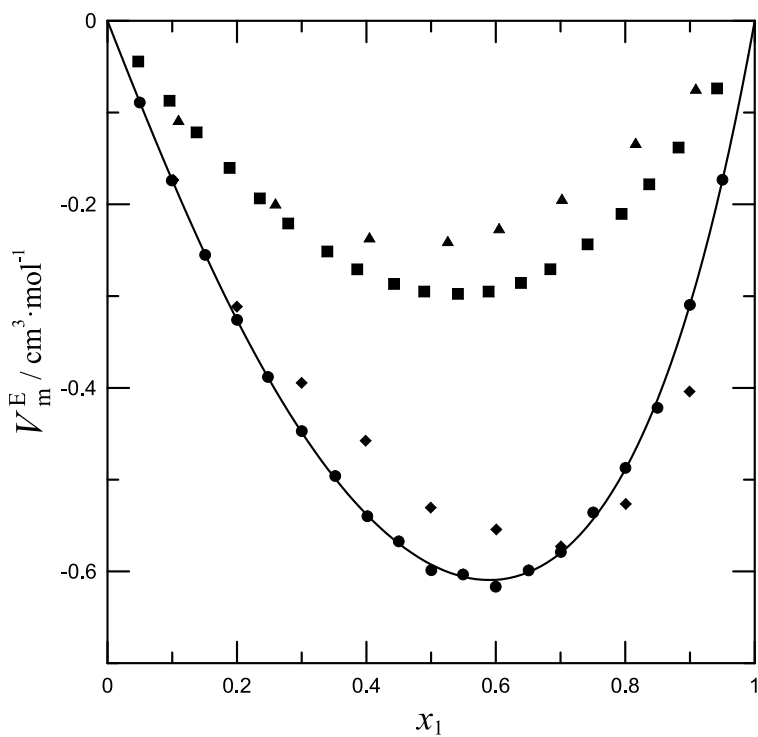

Figure S2. Excess molar volume ($V_m^E$) of the propan-1-ol (1) + aniline (2) liquid mixture as a function of the alkan-1-ol mole fraction ($x_1$) at temperature $T$ and 0.1 MPa. Full symbols, experimental values: (●), this work, $T$ = 298.15 K; (■), ref. [35], $T$ = 298.15 K; (▲), ref. [38], $T$ = 303.15 K; (♦) ref. [31], $T$ = 298.15 K. Solid line, calculations with equation (11) using the coefficients from Table 7.

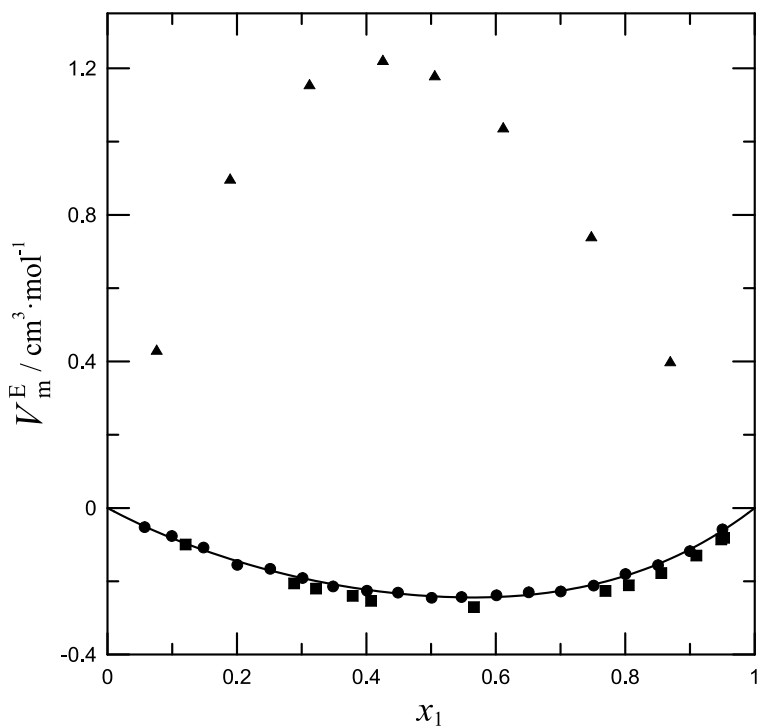

Figure S3. Excess molar volume ($V_m^E$) of the pentan-1-ol (1) + aniline (2) liquid mixture as a function of the alkan-1-ol mole fraction ($x_1$) at temperature $T$ and 0.1 MPa. Full symbols, experimental values: (●), this work, $T$ = 298.15 K; (■), ref. [30], $T$ = 293.15 K; (▲), ref. [38], $T$ = 303.15 K. Solid line, calculations with equation (11) using the coefficients from Table 7.

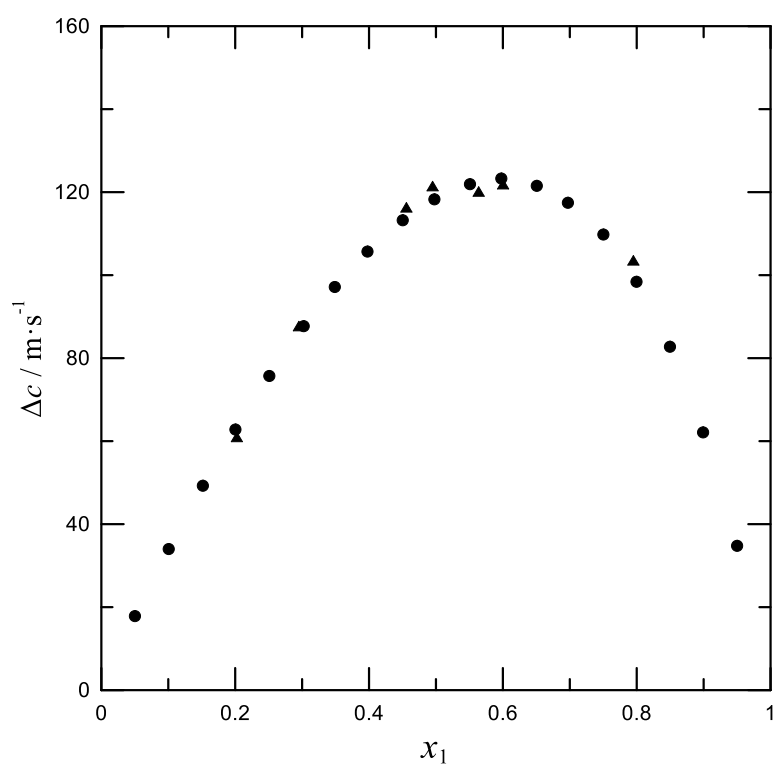

Figure S4. Deviation from linearity of speed of sound ($\Delta c$) of the methanol (1) + aniline (2) liquid mixture as a function of the alkan-1-ol mole fraction ($x_1$) at temperature $T$ and 0.1 MPa. Full symbols, experimental values: (●), this work, $T = 298.15$ K; (▲), ref. [22], $T = 308.15$ K.

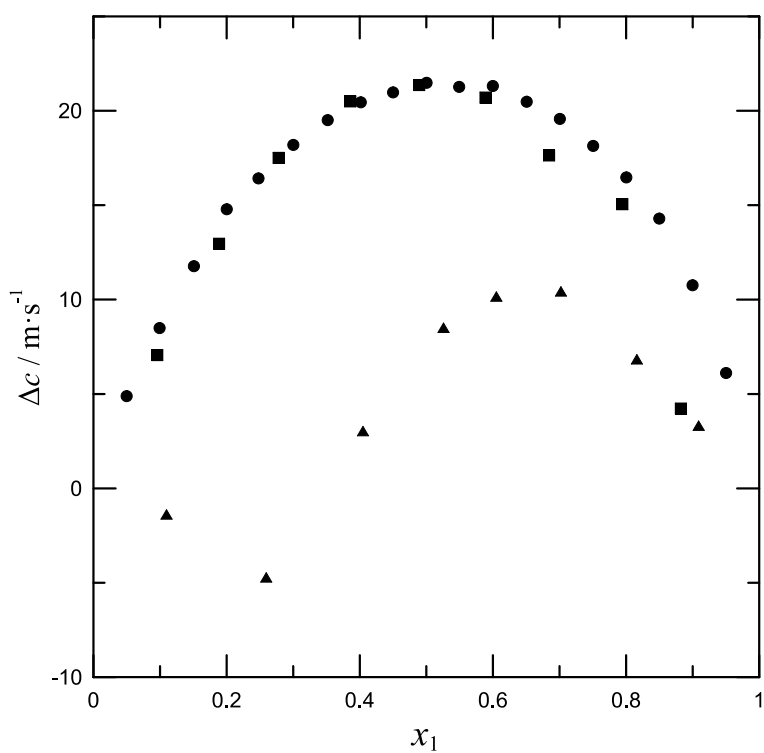

Figure S5. Deviation from linearity of speed of sound ($\Delta c$) of the propan-1-ol (1) + aniline (2) liquid mixture as a function of the alkan-1-ol mole fraction ($x_1$) at temperature $T$ and 0.1 MPa. Full symbols, experimental values: (●), this work, $T$ = 298.15 K; (■), ref. [36], $T$ = 298.15 K; (▲), ref. [38], $T$ = 303.15 K.

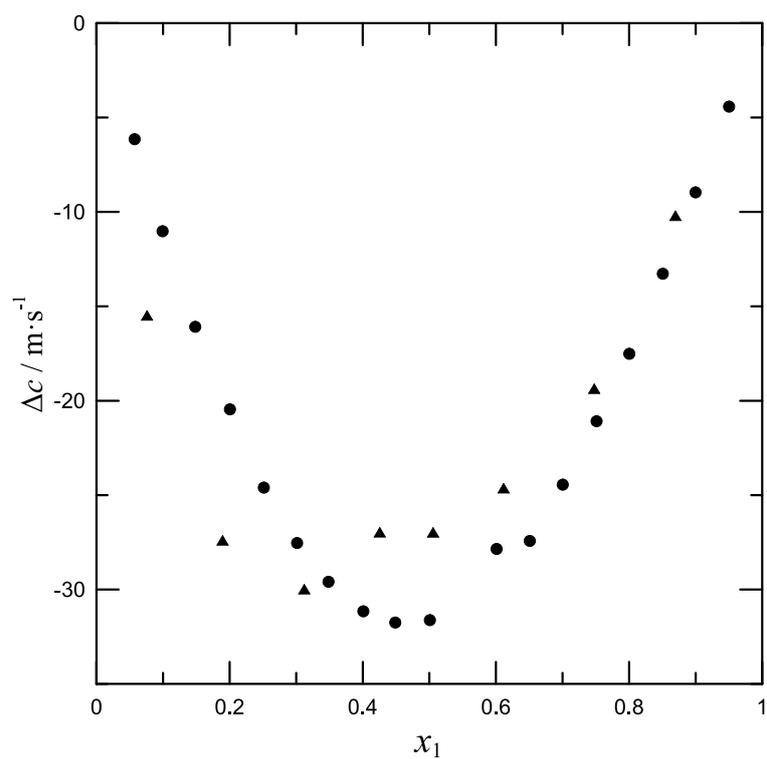

Figure S6. Deviation from linearity of speed of sound ($\Delta c$) of the pentan-1-ol (1) + aniline (2) liquid mixture as a function of the alkan-1-ol mole fraction ($x_1$) at temperature $T$ and 0.1 MPa. Full symbols, experimental values: (●), this work, $T = 298.15$ K; (▲), ref. [38], $T = 303.15$ K.

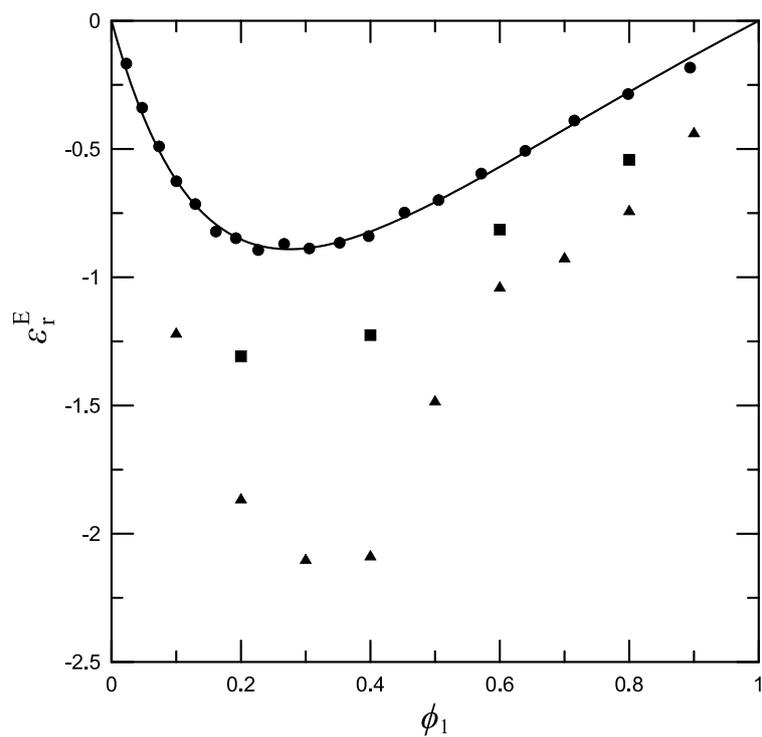

Figure S7. Excess relative permittivity ($\varepsilon_r^E$) of the methanol (1) + aniline (2) liquid mixture as a function of the alkan-1-ol volume fraction ($\phi_1$) at 293.15 K, 0.1 MPa, and 1 MHz. Full symbols, experimental values: (●), this work; (■), ref. [20]; (▲), ref. [26]. Solid line, calculations with equation (11) using the coefficients from Table 7.

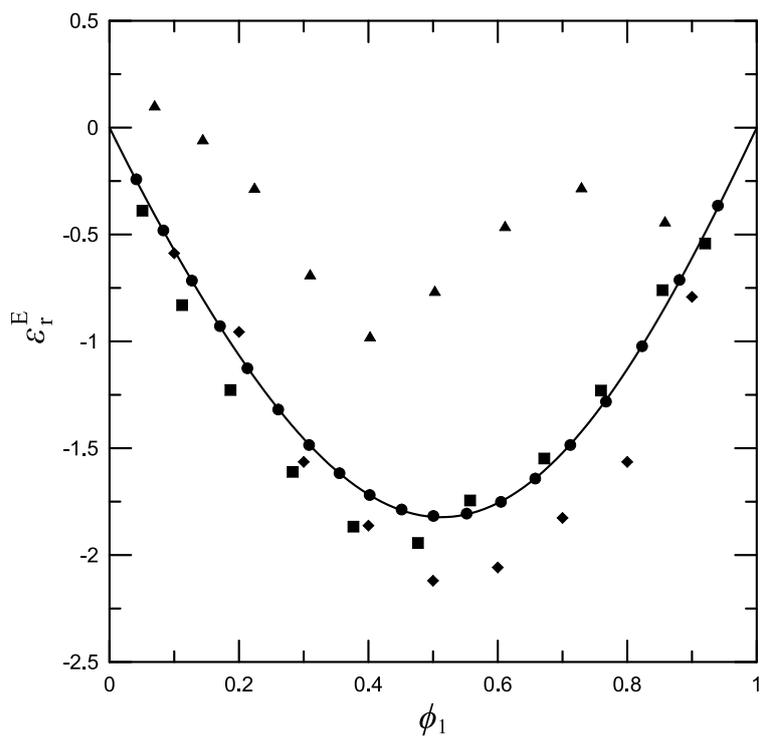

Figure S8. Excess relative permittivity ($\varepsilon_r^E$) of the propan-1-ol (1) + aniline (2) liquid mixture as a function of the alkan-1-ol volume fraction ($\phi_1$) at temperature $T$, 0.1 MPa, and 1 MHz. Full symbols, experimental values: (●), this work, $T$ = 303.15 K; (■), ref. [24], $T$ = 307.15 K; (▲), ref. [28], $T$ = 303.15 K; (♦) ref. [32], $T$ = 303.15 K. Solid line, calculations with equation (11) using the coefficients from Table 7.

# References

See reference list in the main body of the article.